\documentclass[10pt,aps,prd,twocolumn,showpacs,superscriptaddress,nofootinbib,nobibnotes,longbibliography,floatfix]{revtex4-2}

\usepackage{multirow}
\usepackage{bm}
\usepackage{mathtools}
\usepackage{amsmath}
\usepackage{amssymb}
\usepackage{amsfonts}
\usepackage{mathrsfs}
\usepackage{graphicx}
\usepackage{tensor}
\usepackage{accents}
\usepackage{commath}
\usepackage{chngcntr}
\usepackage{siunitx}
\usepackage[dvipsnames]{xcolor}
\usepackage{hyperref}
\hypersetup{colorlinks=true, citecolor=MidnightBlue,
            linkcolor=MidnightBlue, urlcolor=MidnightBlue, linktocpage=true}
\usepackage[normalem]{ulem}
\usepackage{journals}

\usepackage{orcidlink}
\usepackage{booktabs}
\newcommand{\ra}[1]{\renewcommand{\arraystretch}{#1}} 
\newcommand{\orcid}[1]{\href{https://orcid.org/#1}{\includegraphics[width=10pt]{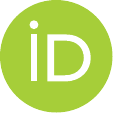}}}

\newcommand{\etal}{\emph{et al.}}

\begin{document}

\title{Probing Dark Star Parameters Through $f$-Mode Gravitational Wave Signals}

\author{Mariachiara Celato \orcid{0009-0003-7478-2922}}
    \email{mariachiara.celato@uni-tuebingen.de}
    \affiliation{Theoretical Astrophysics, IAAT, University of T\"ubingen, 72076 T\"ubingen, Germany}
\author{Christian J. Kr\"uger \orcid{0000-0003-2672-2055}}
    \email{christian.krueger@tat.uni-tuebingen.de}
    \affiliation{Theoretical Astrophysics, IAAT, University of T\"ubingen, 72076 T\"ubingen, Germany}
\author{Kostas D. Kokkotas \orcid{0000-0001-6048-2919}}
    \email{kostas.kokkotas@uni-tuebingen.de}
    \affiliation{Theoretical Astrophysics, IAAT, University of T\"ubingen, 72076 T\"ubingen, Germany}
    
\date{\today}

\begin{abstract}
Theoretical models of self-interacting dark matter offer a promising solution to several unresolved issues within the
collisionless cold dark matter framework.
For asymmetric dark matter, these self-interactions may encourage
gravitational collapse, potentially leading to the creation of compact objects primarily composed of dark matter.
By considering both fermionic and bosonic equations of state, we analyze the equilibrium structure of
non-rotating dark stars, examining their bulk properties and comparing them with baryonic neutron stars.
We show that the frequency and damping rate of $f$-mode oscillations of dark compact stars can be expressed in
terms of universal functions of stellar mass, radius and moment of inertia.
Finally, by employing the universality in the $f$-mode, we propose a scheme to infer accurate values of the physical
parameters of dark compact stars from their $f$-mode gravitational wave signal.
\end{abstract}

\maketitle

\section{Introduction}

The possibility that dark matter (DM) exists as particles that experience significant self-interactions is a
compelling hypothesis, supported by both theoretical considerations and observational evidence.
Theoretically, if the dark sector is part of a unified framework in theories extending beyond the Standard Model,
it is plausible that DM particles interact among themselves through certain gauge bosons.
Additionally, DM self-interactions might be necessary given that the current collisionless cold dark matter (CCDM)
paradigm appears to be inconsistent with some observations.

There are three primary challenges facing the CCDM model.
Firstly, the DM density profile at the centers of dwarf galaxies tends to be flat, whereas CCDM simulations predict
a cuspy profile~\cite{moore, Navarro_1997}.
Observations of rotation curves in these galaxies suggest a discrepancy.
Secondly, CCDM simulations forecast a greater number of satellite galaxies around the Milky Way than what is observed,
as pointed out in Refs.~\cite{Klypin_1999,moore, 10.1093}.
Thirdly, there is the \textit{too big to fail} problem, where simulations predict massive dwarf galaxies that should
have formed observable stars, yet these galaxies are not detected \cite{10.1111}.

In Refs.~\cite{Oh_2011, governato, pontzen}, the authors explore how
these discrepancies might be mitigated by considering DM-baryon interactions or statistical anomalies in
the Milky Way's satellite population.
Another plausible solution is substantial DM self-interactions, which could address all three problems
\cite{10.1111/j.1365-2966.2012.21182.x, Rocha, zavala}.
For instance, DM self-interactions could enhance self-scattering rates in regions with high DM density,
leading to flattened density profiles in dwarf galaxy cores.

The study of DM self-interactions has been extensive, with research exploring various contexts
(cf., e.g., Refs.~\cite{Rocha, zavala, peter, PhysRevLett.84.3760, 2023ResPh..5306967Z}).
Generally, it is believed that DM self-interaction cross-sections within the range of
0.1 cm$^2$/g $< \sigma_{\textrm{XX}}/m_{\textrm{X}} <$ 10 cm$^2$/g (where $\sigma_{\textrm{XX}}$ is the DM self-interaction cross section and $m_{\textrm{X}}$
is the DM particle mass) are sufficient to resolve the challenges faced by the CCDM paradigm.

DM self-interactions can be a valuable feature, potentially aiding the clumping of DM and the formation
of compact objects, especially if these interactions are dissipative or accelerate gravothermal evolution~\cite{balberg}.
Two primary scenarios arise based on the nature of DM annihilation events: one involves DM with significant
annihilation events, and the other involves DM with minimal annihilation events.
The weakly interacting massive particle (WIMP) paradigm fits into the first category.
In this model, the DM relic density in the Universe is determined by DM annihilations.
Initially, DM is in thermal equilibrium with the primordial plasma until the rate of annihilations falls
below the Universe's expansion rate.
In the WIMP framework, DM particles and antiparticles are present in equal numbers.
Gravitational collapse in this scenario could lead to the formation of dark stars, which, according to
Refs.~\cite{PhysRevLett.100.051101, Freese_2009, Freese_2008}, counteract further
collapse through radiation pressure.
However, such stars likely no longer exist because DM annihilations would have
significantly reduced the DM population, causing these dark stars to vanish over time.

Conversely, asymmetric DM presents a viable pathway for forming compact, stable starlike objects.
This scenario serves as a compelling alternative to the WIMP paradigm, and it is well motivated~\cite{NUSSINOV198555, BARR1990387, PhysRevD.74.095008,
PhysRevD.80.037702, PhysRevD.75.085018, PhysRevD.78.115010,
PhysRevD.79.015016, PhysRevD.79.115016, PhysRevD.81.097704}.
In asymmetric DM models, a conserved quantum number, analogous to the baryon number, plays a crucial role.
A mechanism similar to that which generated the baryon asymmetry in the Universe could also produce a
particle-antiparticle asymmetry in the DM sector.
Annihilations in this scenario would eliminate the species with fewer numbers,
leaving an excess population that accounts for the DM relic density.
In the asymmetric DM framework, substantial annihilations do not occur today due to the absence of DM antiparticles.
Therefore, if DM self-interactions can promote collapse, asymmetric DM could form stable compact objects that might
still be detectable today.

The potential for asymmetric DM to form compact objects has been explored for both fermionic and
bosonic DM, for example in Refs.~\cite{PhysRevD.74.063003, PhysRevD.92.063526, 2016JHEP...02..028E}.
These papers have investigated the mass-radius relations, density profiles, and maximum \textit{Chandrasekhar}
mass limits across a broad spectrum of DM particle masses and self-coupling strengths, considering both attractive
and repulsive interactions.
An equation of state (EoS) for asymmetric dark matter, incorporating both
attractive and repulsive interactions, has also been constructed in
Ref.~\cite{PhysRevD.99.083008}. In that work, equilibrium sequences and
maximum masses of asymmetric dark matter stars are identified, and the
stability of neutron stars containing fermionic asymmetric dark matter with an
attractive force is evaluated.
Additionally, Maselli $\etal$~\cite{2017PhRvD..96b3005M} investigate the structure of slowly rotating and tidally
deformed stars, modeled with a DM equation of state based on fermionic and bosonic DM particles, and they show that these
dark objects admit the $I$-Love-$Q$ universal relations, which link their moment of inertia, tidal deformabilities and
quadrupole moments.

In particular, tidal effects on binary stars are measurable in the late
inspiral stage \cite{PhysRevD.77.021502, 2008ApJ...677.1216H,
2009ApJ...697..964H}. Further, 
the measurement of tidal deformations can be used to
extract information about the internal structure of NSs
\cite{PhysRevD.84.024017, PhysRevD.83.084051, PhysRevLett.116.181101,
PhysRevC.95.015801}. Wahidin and Sulaksono study the tidal
deformation of binary dark stars composed of fermionic dark matter, exploring
the effect of dark particle mass and scalar boson coupling exchange on the
EoS, mass-radius relation and tidal Love number properties~\cite{Wahidin_2019}.

The effects of dark matter on the tidal deformability have been explored
also in the case of admixed neutron stars. The presence of a DM core with a
mass fraction ~5\% could affect significantly the interpretation of NS data as
constraints on the nuclear EoS \cite{PhysRevD.97.123007}. A more recent work
on the tidal deformability of dark matter admixed neutron stars can be
found in Ref.~\cite{PhysRevD.105.123010}. 
The impact of asymmetric fermionic dark matter on the thermal evolution of
dark matter admixed neutron stars is discussed in Ref.~\cite{particles7010010}.

Ellis $\etal$ \cite{ELLIS2018607} analyse the impact of dark matter on gravitational
wave signals from a binary merger and find that dark matter cores in merging neutron stars
may yield an observable supplementary peak in the gravitational wave power spectral density following the merger.

In astrophysics, it is customary to develop models of compact stars and establish correlations among various
physical quantities deducible from astronomical observations which are
known as \emph{universal relations} or \emph{asteroseismological relations}
(in case they contain seismological quantities such as mode frequency and/or
damping time).
This approach is motivated by the potential to constrain the EoS for nuclear matter and extract physical parameters of
compact stars. Andersson and Kokkotas~\cite{1998MNRAS.299.1059A} published the
seminal paper opening up the field of gravitational wave
asteroseismology in which they propose the first asteroseismological relations for
oscillation modes of neutron stars. Similar universal relations involving the
$f$-mode frequency and damping time have been proposed subsequently
\cite{Lau_2010, 2015PhRvD..91d4034C, 2018GReGr..50...12L}, some of which also were concerned with
rotating neutron stars~\cite{2008PhRvD..78f4063G, 2015PhRvD..92l4004D,
2021PhRvD.104d3011L, 2020PhRvL.125k1106K}.
Numerous universal relations that correlate bulk quantities other than mode properties
have also been discovered, see, e.g., Refs.~\cite{1996ApJ...456..300L,
2005ApJ...629..979L, 2013Sci...341..365Y, 2016MNRAS.459..646B,
2024PhRvD.109j3033M, 2022ApJ...934..139K, 2023PhRvD.108l4056K}.

There are several avenues for measuring $f$-mode characteristics in
compact stars. For example, it is expected that third-generation detectors
such as Einstein Telescope \cite{2010CQGra..27s4002P} or Cosmic Explorer
\cite{2017CQGra..34d4001A} will have the necessary sensitiviy for such
observations \cite{2012PhRvD..85b4030Z, 2020NatCo..11.2553P}. At the current
time, an estimate for the $f$-mode frequency may be obtained via the $f$-Love
relations \cite{2014PhRvD..90l4023C}, as the tidal deformability $\Lambda$ is
accessible during the late inspiral of a binary system
\cite{2013PhRvD..88d4042R, 2018PhRvL.121i1102D}. Furthermore, the $f$-mode
frequency and damping time can be linked to the moment of inertia $I$; this
quantity may be measurable, e.g., by analyzing spin-orbit coupling effects in
double pulsar systems \cite{2005ApJ...629..979L, 2024APh...15802935A}.

In this paper we present previously proposed relations for NSs involving the
$f$-mode characteristics, which have been reconsidered in
different contexts due to being fundamental to the field
\cite{1998MNRAS.299.1059A, Lau_2010, 2015PhRvD..91d4034C, 2019JCAP...06..051V}, and we investigate
the accuracy of these relations for
isolated stars composed of DM.
We consider static, spherically symmetric objects, modeled with a DM EoS based on fermionic and bosonic DM particles.
We show that these dark stars do not follow the same universal relations that hold true for a large variety of NS EoSs,
and consequently we propose new universal relations for dark stars.
Based on these universal relations, we propose a scheme to determine the mass, radius and moment of inertia of a dark
compact star once its $f$-mode frequency is found from GW observations.

The paper is organized as follows: In Sec.~\ref{DMEoS} we describe the main properties of the two classes of EoSs
considered, and we analyze the bulk properties of the stellar models, such as masses and radii, through their
mass-radius profiles.
In Sec.~\ref{FreqUR} we investigate the accuracy of previously proposed relations between stellar pulsation frequencies
and stellar parameters, and we propose new universal relations for dark stars.
In Sec.~\ref{DampingUR} we do the same as in the previous section, but for relations involving stellar pulsation damping
times and stellar parameters.
In Sec.~\ref{sec:love_numbers} we consider the two universal relations proposed by Chan $\etal$~\cite{2014PhRvD..90l4023C}
and Kuan $\etal$~\cite{2022MNRAS.513.4045K}, which relate the $f$-mode frequency to the tidal deformability $\Lambda$,
testing their applicability to DM stars and, consequently, proposing alternative relations for these models.
In Sec.~\ref{sec:inversion-scheme} we suggest a feasible scheme to apply our findings reported in
Sec.~\ref{sec:universal-relations} to estimate the mass, radius and moment of inertia of a dark compact star if its
$f$-mode GW signal is detected.
We finally summarize our paper in Sec.~\ref{sec:conclusions} and provide some conclusions.
Throughout the whole work we use units in which $c = G = 1$, unless otherwise specified.

\section{Dark Matter Equation of State}\label{DMEoS}

In this section, we describe the most important features of the dark matter EoS
used in this paper to model fermion and boson stars.
In particular, we adopt the same EoSs considered by Maselli\,\etal\,\cite{2017PhRvD..96b3005M}.
\begin{figure}
    \centering
    \includegraphics[width=0.5\textwidth]{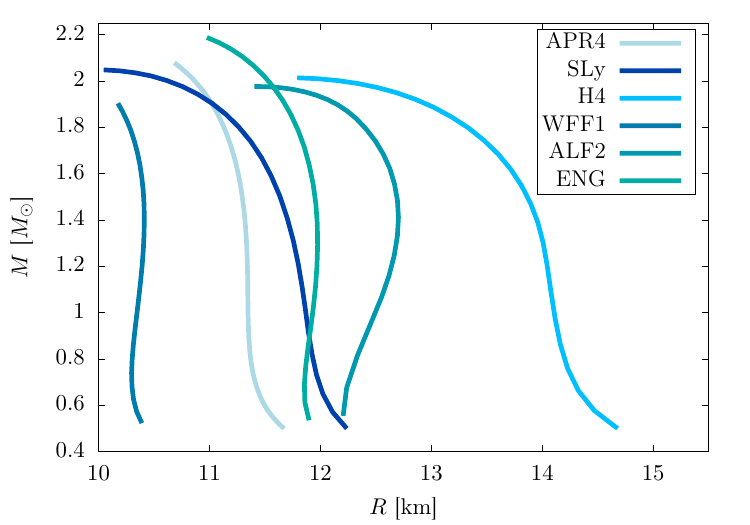}~\caption{Mass-radius profiles for the neutron star EoSs
    used in this work.
    The curves start at $M=0.5\,M_\odot$, and extend to the last stable
    or causal model.
    For some EoSs, the maximal mass model is not determined by
    stability, i.e. $\frac{\partial M}{\partial \epsilon_c} = 0$, but by
    causality as the speed of sounds becomes superluminal already at lower
    $\epsilon_c$ (in our case APR4, ENG, and WFF1).
    }
    \label{fig-mr-ns}
\end{figure}

Moreover, in our analysis we will also consider the piecewise polytropic
approximations \cite{2009PhRvD..79l4032R} to 6 neutron star EoSs
(APR4 \cite{PhysRevC.58.1804},
SLy \cite{refId0},
H4 \cite{PhysRevD.73.024021},
WFF1 \cite{PhysRevC.38.1010},
ALF2 \cite{Alford_2005},
ENG \cite{1996ApJ...469..794E}), which will allow one to make a direct comparison
between the macroscopic features of baryonic and dark objects.
The mass-radius profiles obtained by using the aforementioned EoSs are shown in Fig.~\ref{fig-mr-ns}.
The curves start from light stars with a mass of $M=0.5\,M_{\odot}$, and extend to the last stable model.
However, with some EoSs, the curves do not reach the neutron star model
for which $\frac{\partial M}{\partial \epsilon_c} = 0$, because the EoS
becomes acausal already at lower energy densities. We show the
pressure vs. energy density relations of these EOSs in Figs.~\ref{fig-eos-fdm}
and \ref{fig-eos-bdm} along with the dark matter relations; see the following
two subsections.

\subsection{Fermion star}

We consider a fermionic particle interacting via a repulsive Yukawa potential, e.g., due to a massive dark photon~\cite{PhysRevD.92.063526}
\begin{equation}
    V(r) = \alpha_{\textrm{X}} \frac{e^{-\hbar m_{\phi} r}}{r},
\end{equation}
where $\alpha_{\textrm{X}}$ is the dark fine structure constant, $\hbar$ the
reduced Planck constant and $m_{\phi}$ is the mass of the mediator.

Models incorporating a Yukawa potential are particularly valuable in the study of self-interacting DM,
as they effectively suppress the scattering cross section at high relative velocities, as shown in Ref.~\cite{PhysRevD.87.115007}.
This characteristic preserves the viability of collisionless cold DM on supergalactic scales while simultaneously
addressing the flattening of subgalactic structures.

\begin{figure}
    \centering
    \includegraphics[width=0.5\textwidth]{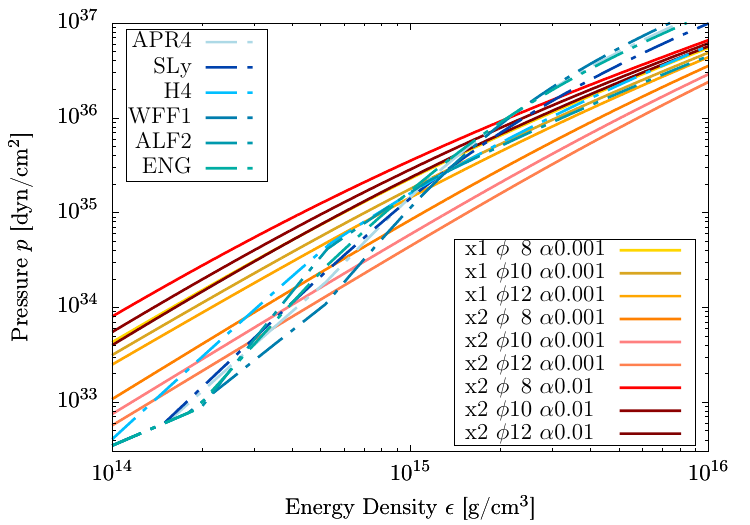}
    \caption{Relation between pressure $p$ and energy density $\epsilon$ for
    the fermionic dark matter EOSs employed in this study, which are shown in
    different shades of yellow and red. For comparison, we also show the
    relations of six nuclear matter EOSs in shades of blue with dashed lines.}
    \label{fig-eos-fdm}
\end{figure}

Pressure in fermion stars has two contributions: one from Fermi repulsion and one due to the Yukawa interactions.
By calculating the energy density and pressure due to Yukawa interactions in the mean field approximation, the EoS is given by two
implicitly related equations:
\begin{align}
    &\epsilon(x) = \frac{m_{\textrm{X}}^4 c^3}{\hbar^3}\left[\xi(x)+  \frac{2}{9\pi^3}\frac{\alpha_{\textrm{X}}}{\hbar c}\frac{m_{\textrm{X}}^2}{m_{\phi}^2}x^6\right],
    \label{eps_eos_fermions} \\
    &P(x) = \frac{m_{\textrm{X}}^4 c^5}{\hbar^3}\left[\chi(x)+ \frac{2}{9\pi^3}\frac{\alpha_{\textrm{X}}}{\hbar c}\frac{m_{\textrm{X}}^2}{m_{\phi}^2}x^6\right],
    \label{p_eos_fermions}
\end{align}
where $x\equiv p_{\textrm{F}}/(m_{\textrm{X}} c)$ is a dimensionless quantity that measures the Fermi momentum compared to the DM mass.
The functions $\xi$ and $\chi$ are the contributions from Fermi repulsion, given by
\begin{align*}
    &\xi(x) = \frac{1}{8\pi^2}\left[x\sqrt{1+x^2}(2x^2+1)-\ln\left(x+\sqrt{1+x^2}\right)\right], \\
    &\chi(x) = \frac{1}{8\pi^2}\left[x\sqrt{1+x^2}(2x^2/3-1)+\ln\left(x+\sqrt{1+x^2}\right)\right].
\end{align*}
Both pressure and density are smooth monotonic functions of the parameter $x$.
We show the resulting pressure vs. energy density relations in
Fig.~\ref{fig-eos-fdm}.

\begin{figure}
    \centering
    \includegraphics[width=0.5\textwidth]{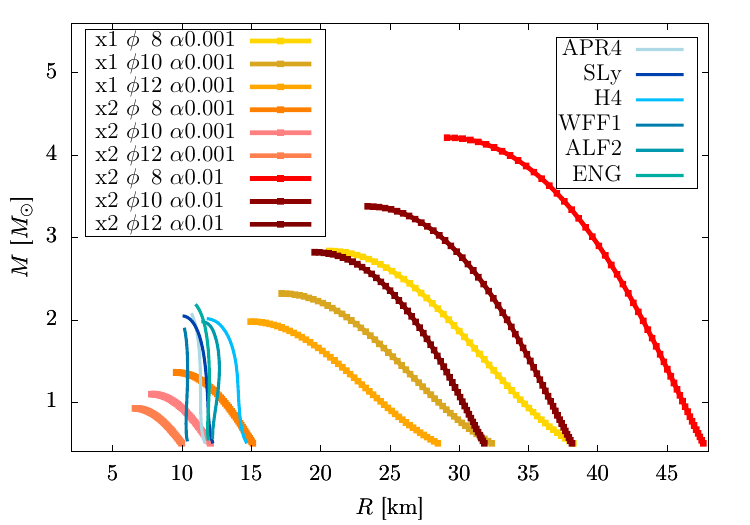}~\caption{Mass-radius profiles for fermionic dark matter stars
    with $\alpha$ = (10$^{-3}$, 10$^{-2}$), dark particle mass $m_{\textrm{X}}$ = (1,2) GeV, and mediator mass
        $m_{\phi}$ = (8, 10,12) MeV.
    The curves start at $M=0.5\,M_\odot$, and extend to the last stable model; they are shown in different shades of yellow and red. For comparison, we also show the mass-radius curves of six nuclear matter EoS in shades of blue.}
    \label{fig-mr-fdm}
\end{figure}
Hereafter, we will vary the three parameters that fully specify fermion stars in the ranges
$\alpha_{\textrm{X}}/(\hbar c) \equiv \alpha = (10^{-3}, 10^{-2})$, $m_{\phi} = (8, 10, 12)$ MeV and $m_{\textrm{X}}= (1,2)$ GeV.
The mass-radius relations for all our models are presented in
Fig.~\ref{fig-mr-fdm}. We use a specific identifier for fermionic EoS that
contains the three parameters $m_\textrm{X}$, $m_\phi$, and $\alpha$.
Exemplarily, the EoS specified by the parameter set $m_\textrm{X} = 1\,$GeV, $m_\phi = 10\,$MeV and $\alpha = 10^{-2}$ will be labeled as ``x1 $\phi$10 $\alpha$0.01''.
Each point of the plot is obtained, for a given EoS, by varying the star's central pressure and solving the well-known
Tolman-Oppenheimer-Volkoff equations with that value of the central pressure and the corresponding central energy density.
It can be seen that, for a fixed value of the radius, increasing values of $m_{\textrm{X}}$ significantly reduce the mass,
resulting in less compact objects.
Similarly, variations in the mediator mass $m_{\phi}$ yield substantial changes in the same direction as $m_{\textrm{X}}$.
In the same plot, we include also the mass-radius profiles of the baryonic EoS considered in this study.
We note that there are overlapping regions between fermion stars and baryonic matter profiles, suggesting the potential
for similar configurations.
However, baryonic EoSs typically exhibit steeper slopes, resulting in greater variations of the compacteness
$\textit{C} = M / R$ of the corresponding neutron star models.

\subsection{Boson star}

Boson stars are inherently smaller than their fermionic counterparts because they lack Fermi pressure to
counterbalance their self-gravity.
In the absence of self-interactions, boson stars rely on quantum mechanical pressure arising from the uncertainty
principle for stabilization.
However, unless the bosons are exceptionally lightweight, this pressure is naturally minimal and can only
support small concentrations of matter.
Introducing self-interactions would allow a boson star to achieve a size comparable to that of a fermion star.

\begin{figure}
    \centering
    \includegraphics[width=0.5\textwidth]{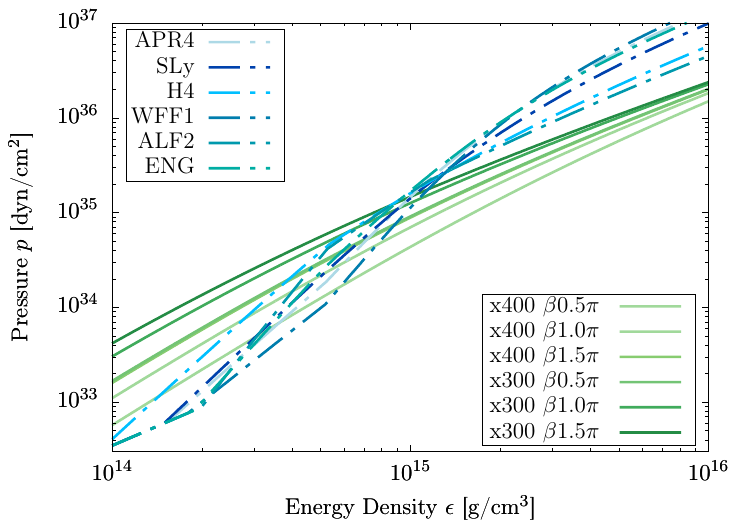}
    \caption{Same as Fig.~\ref{fig-eos-fdm} but for the bosonic dark
    matter EoSs.}
    \label{fig-eos-bdm}
\end{figure}
\begin{figure}
    \centering
    \includegraphics[width=0.5\textwidth]{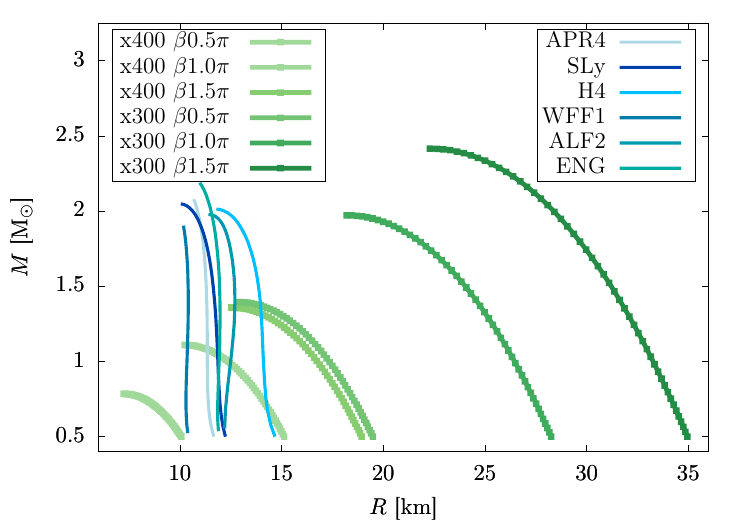}~\caption{Mass-radius profiles for bosonic dark matter stars
    with different values of the coupling constant $\beta=(0.5,1.0,1.5)\pi$ and boson mass $m_{\textrm{X}}=(300,400)$ MeV.
    The curves start at $M=0.5\,M_\odot$, and extend to the last stable model. For comparison, we also show the mass-radius curves of six nuclear matter EOS in shades of blue.}
    \label{fig-mr-bdm}
\end{figure}

In this work, we consider boson star models whose EoS was first derived in Ref.~\cite{PhysRevLett.57.2485}, and it is given by
\begin{equation}
    P = \frac{c^5}{9\hbar^3}\frac{m_{\textrm{X}}^4}{\beta} \left( \sqrt{1 + \frac{3\hbar ^3}{c^3} \frac{\beta \epsilon}{m_{\textrm{X}}^4}} -1 \right) ^2,
    \label{eos_bdm}
\end{equation}
where $m_{\textrm{X}}$ is the boson mass and $\beta$ is a dimensionless coupling constant.
We show the pressure vs. energy density graphs of the employed bosonic
EoSs in Fig.~\ref{fig-eos-bdm} and the mass-radius profiles for bosonic dark matter stars in Fig.~\ref{fig-mr-bdm}.
For bosonic EoS, we use the following identifier containing $m_\textrm{X}$ and
$\beta$: the EoS specified by the parameter set $m_{\textrm{X}}=400$ MeV,
$\beta = 0.5 \pi$ will be labeled as "x400 $\beta 0.5\pi$".
Two values of the boson mass are considered,
$m_{\textrm{X}} = (300, 400)$ MeV, as well as three different values of the coupling parameter $\beta = (0.5, 1.0, 1.5)\pi$.
It is observed that stronger couplings result in larger radii for a given mass, thereby yielding less compact objects.
A similar trend, albeit more pronounced, is evident when considering lighter dark particles.
Even in the case of boson stars, the curves exhibit a smoother slope compared to those derived for standard nuclear matter.
Finally, we note that stellar models constructed with the bosonic dark matter EoSs considered in this work exhibit
self-similar symmetries.
Namely, the shape of the mass-radius relation is independent of EoS parameters such as the mass of the dark particle
$m_{\textrm{X}}$ or the coupling parameter $\beta$.
A proof of this feature can be found in Ref.~\cite{2017PhRvD..96b3005M} and is also outlined in App.~\ref{sec:self-similarity} of the present
work.

\section{Universal Relations}
\label{sec:universal-relations}

\subsection{Frequency relations of the $f$-mode}\label{FreqUR}
Compact star pulsations can lead to GW emission.
Since GWs carry away energy, they act as a damping mechanism.
In a perturbative approach the pulsations are treated as damped linear oscillations, which are analyzed in terms of quasi-normal
modes (QNMs).
This ansatz assumes an $e^{i\omega t}$ time dependency, where $\omega$ is the
complex-valued eigenfrequency of the QNM.
The complex nature of the eigenfrequency accounts for the damping.
It reads
\begin{equation}
    \omega = 2 \pi f + \frac{i}{\tau},
\end{equation}
where $f$ is the pulsation frequency and $\tau$ the damping time of the oscillation.

We focus on the fundamental $f$-mode whose frequencies we determine using the code presented in Ref.~\cite{ck_phdthesis}.

In the considered case of isolated, static stars, many relations have been
proposed relating the fundamental frequency $f$ to stellar parameters
such as the mass $M$, radius $R$, moment of inertia $I$ and tidal deformability $\Lambda = \frac{2}{3}k_2(\frac{c^2 R}{G M})^5$.
They exhibit a different level of accuracy.

\begin{figure}
    \centering
    \includegraphics[width=0.5\textwidth]{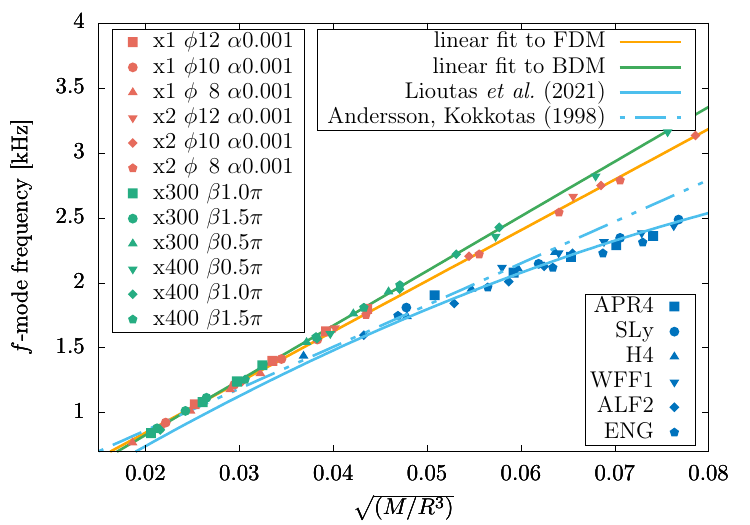}~\caption{
    The numerically obtained $f$-mode frequencies plotted as a function of the mean stellar density.
    Orange symbols represent fermionic dark matter stars with $\alpha = 10^{-3}$, dark particle mass $m_\textrm{X}=(1,2)$ GeV,
    and mediator mass $m_{\phi}=(8,10,12)$ MeV.
    Green symbols stand for bosonic dark matter stars with coupling constant
    $\beta=(0.5,1.0,1.5)\pi$ and boson mass $m_\textrm{X}=(300,400)$ MeV. 
    Light-blue symbols show the behaviour of the NS stellar models chosen for comparison.
    The light-blue dashed-dotted curve corresponds to the universal relation for the $f$-mode frequency of Andersson and Kokkotas~\cite{1998MNRAS.299.1059A} described in Eq.~\eqref{UR_AK_frequency_my},
        while the relation by Lioutas \etal~\cite{2021PhRvD.104d3011L} reported in Eq.~\eqref{UR_LB_frequency_my}
        is shown with a light-blue solid curve.
        Our best linear fits to fermionic (Eq.~\eqref{UR_fdm_myAKstyle})
        and bosonic (Eq.~\eqref{UR_bdm_myAKstyle}) dark matter models are represented by an orange and a green line, respectively.}
    \label{fig_f_AK}
\end{figure}

A very well known relation was proposed by Andersson and Kokkotas~\cite{1998MNRAS.299.1059A} between $f$
and the mean density of the star, and reads
\begin{equation}
    \frac{f}{\textrm{kHz}} \approx 0.22 + 32.16 \sqrt{\frac{M}{R^3}}.
    \label{UR_AK_frequency_my}
\end{equation}
This relation is shown as a light-blue dashed-dotted line in Fig.~\ref{fig_f_AK}.
Years later, Lioutas, Bauswein and Stergioulas proposed a relation also employing the star's average density in Ref.~\cite{2021PhRvD.104d3011L}, but based on a
second-order fit to their data for NSs, namely
\begin{equation}
    \frac{f}{\textrm{kHz}} \approx -0.133 + 47.23 \sqrt{\frac{M}{R^3}} - 173.2 \frac{M}{R^3}.
    \label{UR_LB_frequency_my}
\end{equation}
Equation~\eqref{UR_LB_frequency_my} is represented with a light-blue solid line in Fig.~\ref{fig_f_AK}.
In the same figure, we include some NS data points (light-blue) with the purpose of getting a visual impression of how
the NS data scatter around the corresponding universal relations.
Green points represent instead bosonic dark matter stars obtained by varying the stellar central energy density,
as well as the EoS parameters.
Specifically, the dark boson mass is varied in the range $m_\textrm{X} = (300, 400)$ MeV, while the coupling constant
assumes the values $\beta = (0.5, 1.0, 1.5)\pi$.
Finally, orange data points correspond to fermionic dark matter stars whose EoSs is determined by a dark particle mass
$m_{\textrm{X}} = (1.0, 2.0)$ GeV, a mediator mass $m_{\phi} = (8, 10, 12)$\,MeV and a fixed value of the dark fine structure constant,
set to $\alpha_{\textrm{X}}/(\hbar c) \equiv \alpha = 10^{-3}$.
From these results we can infer that the dark matter EoSs considered in this work do not follow the relation proposed
by Andersson and Kokkotas for NSs, but they seem instead to distribute along straight lines, with two well-defined
slopes, one for bosonic models and one for fermionic ones.
Consequently, we perform a linear fit on our bosonic dark matter's data points, obtaining
\begin{equation}
    \frac{f}{\textrm{kHz}} \approx -0.01461 + 42.11 \sqrt{\frac{M}{R^3}}.
    \label{UR_bdm_myAKstyle}
\end{equation}
As regards fermionic dark matter models, the result of a linear fit is
\begin{equation}
    \frac{f}{\textrm{kHz}} \approx 0.06661 + 38.97 \sqrt{\frac{M}{R^3}}.
    \label{UR_fdm_myAKstyle}
\end{equation}
Both the equations fit the given data points with a relative error smaller than 2\%.
As shown in Table~\ref{tab:relative-error-f-AK}, comparing the values of the $f$-mode frequency with the estimates given
by the universal relations~\eqref{UR_AK_frequency_my} and
~\eqref{UR_LB_frequency_my} for a variable mean stellar density in the range
$\sqrt{M/R^3} \in [0.055,0.075]$,
the DM values deviate from the fit by Andersson and Kokkotas~\cite{1998MNRAS.299.1059A} by more than 10\%, and by more
than 14\% in the case of the fit by Lioutas \etal~\cite{2021PhRvD.104d3011L}.

\begin{table*}
    \centering
    \caption{Relative errors in the estimate of the $f$-mode frequency obtained by using
    universal relations. We specify the dark matter EoS in the first column, the mean stellar
    density $\sqrt{M/R^3}$ in the second, and the numerical value of the $f$-mode frequency in the third.
    Moreover, column four and five correspond to the estimate of $f$ obtained
    by using the universal relation by Anderson and Kokkotas
    of Eq.~\eqref{UR_AK_frequency_my} and its relative error with respect to the true value, respectively.
    The behaviour of the dark matter EoS is not well represented by this universal relation, as the relative error is always
    greater than 10\%.
    In the last two columns, we show the estimate and relative error for the case of the universal relation by Lioutas \etal~\cite{2021PhRvD.104d3011L} of Eq.~\eqref{UR_LB_frequency_my}. The relative error in the estimate of
    $f$ is grater than 14\%.}
    \ra{1.3}
    \begin{tabular}{@{}ccc|cc|cc@{}}
        \toprule
        EoS & $\sqrt{\frac{M}{R^3}}$ & $f$ & $f_\textrm{AK}$ & $\frac{|f -
        f_\textrm{AK}|}{f}$ & $f_\textrm{LBS}$ & $\frac{|f - f_\textrm{LBS}|}{f}$ \\
        \hline
        x2 $\phi \phantom{0}$8 $\alpha0.001$  & 0.064 & 2.541 & 2.280 & 0.103 & 2.181 & 0.141 \\
        x2 $\phi \phantom{0}$8 $\alpha0.001$  & 0.071 & 2.788 & 2.489 & 0.107 & 2.337 & 0.162 \\
        x2 $\phi$10 $\alpha0.001$ & 0.069 & 2.749 & 2.424 & 0.118 & 2.290 & 0.167 \\
        x2 $\phi$12 $\alpha0.001$ & 0.066 & 2.666 & 2.329 & 0.126 & 2.219 & 0.167 \\
        x400 $\beta 0.5\pi$ & 0.057 & 2.357 & 2.063 & 0.125 & 2.005 & 0.150 \\
        x400 $\beta 0.5\pi$ & 0.068 & 2.821 & 2.406 & 0.147 & 2.277 & 0.193 \\
        x400 $\beta 1.0\pi$ & 0.058 & 2.427 & 2.075 & 0.145 & 2.015 & 0.169 \\
        \bottomrule
    \end{tabular}
    \label{tab:relative-error-f-AK}
\end{table*}

Another relation exploring the connection between the $f$-mode frequency and the star's moment of inertia
was discussed by Lau $\etal$~\cite{Lau_2010}.
In particular, arguing that the moment of inertia $I$ is sensitive to the matter distribution within the star, they
defined an effective compactness $\eta \equiv \sqrt{M^3/I}$.
In this way, they accounted for the different density profiles of neutron stars and quark stars, which prevented the
previously suggested relations from being applicable to quark stars.
The relation proposed by Lau $\etal$ is a quadratic expression in $\eta$ of the form:
\begin{equation}
    M \omega_r \approx -0.0047 + 0.133 \eta + 0.575 \eta^2,
    \label{UR_Lau_freq}
\end{equation}
where $\omega_r \equiv 2\pi f$.
We note that Eq.~\eqref{UR_Lau_freq} is slightly different from the corresponding Eq. (6) of Ref.~\cite{Lau_2010},
as the latter contained a typo in the coefficient of the quadratic term.
\begin{figure}
    \centering
    \includegraphics[width=0.5\textwidth]{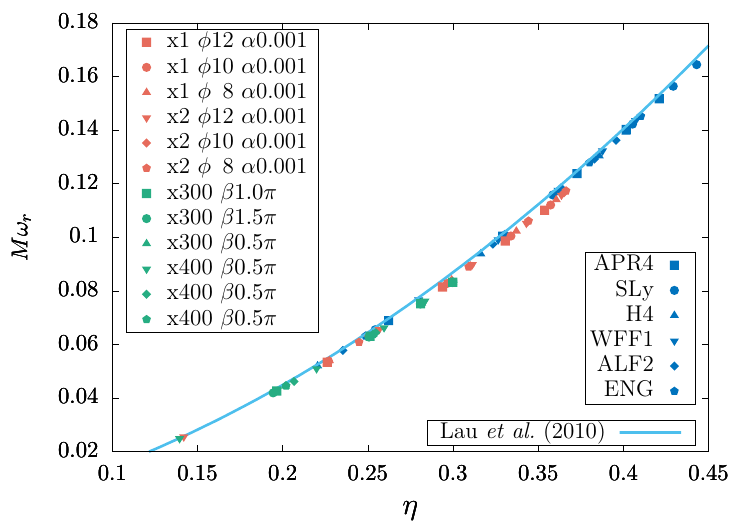}
    \caption{The mass-scaled real part of the $f$-mode frequency $\omega_r = 2\pi f$ is plotted against the
    effective compactness $\eta = \sqrt{M^3 / I}$.
        The blue curve represents the universal relation proposed by Lau~\etal~\cite{Lau_2010}
    and correponds to Eq.~\eqref{UR_Lau_freq}. Green, orange and blue points correspond to stellar models as described in
    Fig.~\ref{fig_f_AK}.}
    \label{fig:UR_Lau_f}
\end{figure}

This relation is shown as a light-blue solid curve in Fig.~\ref{fig:UR_Lau_f}.
Comparing the curve with data points for NSs (light-blue), bosonic DM models (green) and fermionic DM models (orange)
it is clear that this relation fits nicely every type of EoS considered.

\subsection{Damping time relations of the $f$-mode} \label{DampingUR}

\begin{figure}
    \centering
    \includegraphics[width=0.5\textwidth]{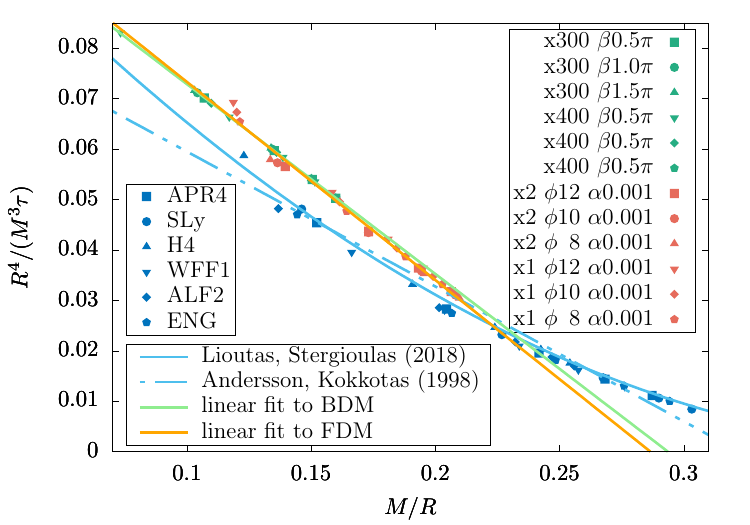}
    ~\caption{
    The normalized damping time of the $f$-mode as a function of the stellar
    compactness.
    The universal relation of Eq.~\eqref{UR_AK_damping_my} proposed by
    Andersson and Kokkotas~\cite{1998MNRAS.299.1059A} is shown as a light-blue dashed-dotted line.
        The blue curve corresponds to Eq.~\eqref{UR_LS_damping_my}, namely to the universal
        relation of Lioutas and Stergioulas~\cite{2018GReGr..50...12L}.
        The best linear fits to our bosonic (cf. Eq.~\eqref{UR_bdm_damping_myAKstyle}) and fermionic
        (cf. Eq.~\eqref{UR_fdm_damping_myAKstyle})
        dark matter stellar models are represented with
        a green and an orange line, respectively.
    Green, orange and blue points represent stellar models as described in
    Fig.~\ref{fig_f_AK}.}
    \label{fig:UR_damp_Ak}
\end{figure}

Andersson and Kokkotas~\cite{1998MNRAS.299.1059A} proposed also a universal relation for the damping
time $\tau$ of the $f$-mode.
They estimated the characteristic damping time based on the quadrupole formula, namely
\begin{equation*}
    \tau \sim \frac{\text{oscillation energy}}{\text{power emitted in GW}} \sim \frac{R^4}{M^3},
\end{equation*}
thus finding a relation which reads
\begin{equation}
    \frac{R^4}{M^3 \tau} \approx 0.08627 - 0.2676 \frac{M}{R}.
    \label{UR_AK_damping_my}
\end{equation}
Equation~\eqref{UR_AK_damping_my} is represented as a light-blue dashed-dotted line in Fig.~\ref{fig:UR_damp_Ak}.

In Ref.~\cite{2018GReGr..50...12L}, Lioutas and Stergioulas extended the
relation found by Andersson and Kokkotas to higher order.
They found that their numerical data for $\tau$ are described very accurately by the following universal relation
\begin{equation}
    \frac{R^4}{M^3 \tau} \approx 0.112 - 0.53 \frac{M}{R} + 0.628 \left(\frac{M}{R}\right)^2.
    \label{UR_LS_damping_my}
\end{equation}
As can be seen in Fig.~\ref{fig:UR_damp_Ak}, although this quadratic relation in $M/R$ (light-blue solid line)
fits well the neutron stars' points, it fails to represent the behaviour of bosonic (green) and fermionic
(orange) dark matter models, which clearly display, again, a linear distribution.
A linear fit to the bosonic DM models gives

\begin{equation}
    \frac{R^4}{M^3 \tau} \approx 0.1105 - 0.3764 \frac{M}{R},
    \label{UR_bdm_damping_myAKstyle}
\end{equation}
while in the fermionic case we find
\begin{equation}
    \frac{R^4}{M^3 \tau} \approx 0.1126 - 0.3930 \frac{M}{R}.
    \label{UR_fdm_damping_myAKstyle}
\end{equation}
As usual, the two new relations are represented in Fig.~\ref{fig:UR_damp_Ak} via green and orange lines for the
bosonic and fermionic dark matter case, respectively.
The relative error is less than 0.5\% in the former case, and less than 4\% in the latter.

\begin{figure}
    \centering
    \includegraphics[width=0.5\textwidth]{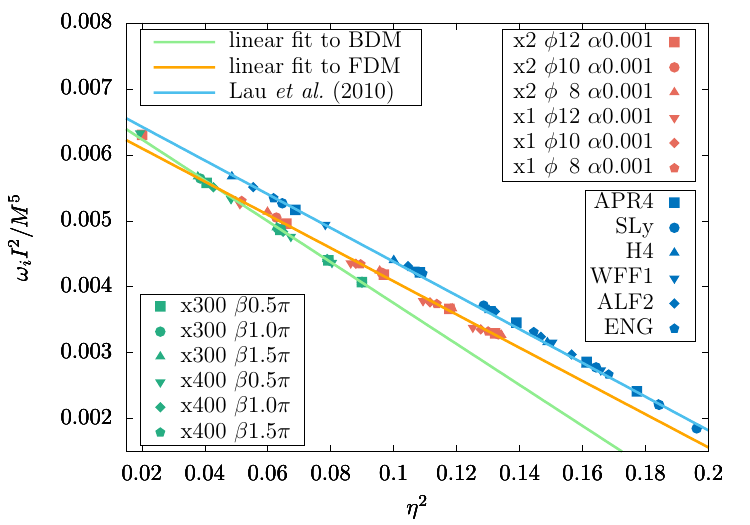}
    ~\caption{
        $\omega_i I^2/M^5$ is plotted against $\eta^2$ (i.e. $M^3/I$).
        The universal relation for the imaginary part of $f$-mode proposed by
        Lau~\etal~\cite{Lau_2010} is shown as
    a blue line (see Eq.\eqref{UR_Lau_damp}).
    The green line represents the best linear fit to our bosonic dark matter models, namely Eq.~\eqref{eq_UR_damp_myLLL_bdm}.
    The orange line is the best linear fit to the fermionic dark matter stars and corresponds to Eq.~\eqref{eq_UR_damp_myLLL_fdm}.
    Green, orange and blue points represent stellar models as described in Fig.~\ref{fig_f_AK}.}
    \label{fig:UR_damp_LLL}
\end{figure}

Another relation for the $f$-mode damping time was proposed by Lau $\etal$~\cite{Lau_2010}.
Their quadratic expression in $\eta$ reads
\begin{equation}
    \frac{I^2}{M^5} \omega_i \approx 0.00694 - 0.0256 \eta^2
    \label{UR_Lau_damp}
\end{equation}
where $\eta \equiv \sqrt{M^3/I}$ and $\omega_i \equiv 1/\tau$.
This relation can be found in Fig.~\ref{fig:UR_damp_LLL} as a light-blue curve, together with NS data points (blue),
bosonic (green) and fermionic (orange) stellar models.
While the data points referring to NSs fall on the aforementioned relation with high accuracy, the dark models seem to
follow different patterns.
More precisely, their distributions suggest that bosonic and fermionic EoSs might be characterised by their own linear
universal relations.
By performing a linear fit to the data, we find that an expression in $\eta$ of the form
\begin{equation}
    \frac{I^2}{M^5}\omega_i \approx 0.006858 -0.03105 \eta^2
    \label{eq_UR_damp_myLLL_bdm}
\end{equation}
yields the best linear fit to the bosonic distribution, while the formula
\begin{equation}
    \frac{I^2}{M^5}\omega_i \approx 0.006604 -0.02521 \eta^2
    \label{eq_UR_damp_myLLL_fdm}
\end{equation}
gives the best linear fit to the fermionic data points.
Eq.~\eqref{eq_UR_damp_myLLL_bdm} and~\eqref{eq_UR_damp_myLLL_fdm} are shown in Fig.~\ref{fig:UR_damp_LLL} with a green
and an orange line, respectively.
We observe that a fit of the linear type works nicely, and with high accuracy.
In particular, the relative error is less than 0.5\% in the bosonic dark matter case, and less than 2\% in the
fermionic one.

\subsection{Tidal Love Numbers}
\label{sec:love_numbers}
The dark stars considered in this study may exist also in binary systems, other than as isolated entities.
These stars might generate GW signals that differ from those produced by black hole binaries.
The EoS leaves a distinctive mark on the signals emitted during binary mergers, primarily influenced by adiabatic tidal
interactions.
These interactions are defined using a series of coefficients known as Love numbers, calculated under the assumption
that tidal effects arise from an external, time-independent gravitational field.
The primary contribution $k_2$, which corresponds to a quadrupolar
deformation, is defined by the following relation \cite{2008ApJ...677.1216H,
2009ApJ...697..964H}
\begin{equation}
    Q_{ij} = \frac{2}{3}k_2 R^5 \xi_{ij} = \Lambda
    \xi_{ij},
\end{equation}
where $\xi_{ij}$ is the external tidal tensor and $Q_{ij}$ is the (tidally deformed) star's quadrupole tensor.
The Love number $k_2$ or, equivalently, the tidal deformability $\Lambda$, depends solely on the star's EoS.

\begin{figure}
    \centering
    \includegraphics[width=0.5\textwidth]{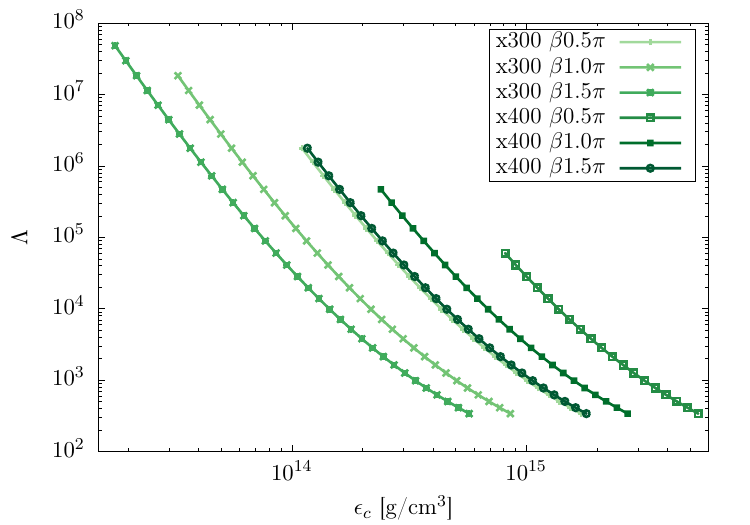}
    \caption{Tidal deformability $\Lambda$ vs. central energy density $\epsilon_c$ for the bosonic dark matter EoSs
    considered in this study. The EoSs are labelled according to the value of their coupling constant $\beta=(0.5, 1.0, 1.5)\pi$
    and boson mass $m_{\textrm{X}}=(300,400)$ MeV.
    For each curve, the lowest value of $\Lambda$ corresponds to the tidal deformability of the last stable model, while
the highest value identifies a stellar model with mass equal to
$M=0.5\,M_{\odot}$.}
    \label{fig:lambda_vs_eps}
\end{figure}

Figure~\ref{fig:lambda_vs_eps} shows the behaviour of the tidal deformability $\Lambda$ as a function of the central energy
density $\epsilon_c$ for the bosonic dark matter EoS considered in this study.
For each curve, the lowest value of $\Lambda$ corresponds to the tidal deformability of the last stable model, while
the highest value identifies a stellar model with mass equal to $M=0.5\,M_{\odot}$.
It can be seen that $\Lambda$ decreases a lot as the central energy density of the star is increased.
For a given value of $\epsilon_c$, the tidal deformability $\Lambda$ decreases as the dark matter coupling constant $\beta$ is increased, while bigger
values of the dark particle mass $m_{\textrm{X}}$ give the opposite effect.

Universal relations exploring the connection between the $f$-mode frequency $f$ and the tidal deformability $\Lambda$
have been proposed by Chan $\etal$~\cite{2014PhRvD..90l4023C} and also by Kuan~$\etal$~\cite{2022MNRAS.513.4045K}.
In order to investigate the validity of those universal relations for the case of bosonic dark matter,
we select values of $\Lambda$ in the range from 400 to 15000.
Indeed, within this range, each bosonic dark matter EoS can provide stable stellar models heavier than 0.5 $M_{\odot}$.

\begin{figure}
    \centering
    \includegraphics[width=0.5\textwidth]{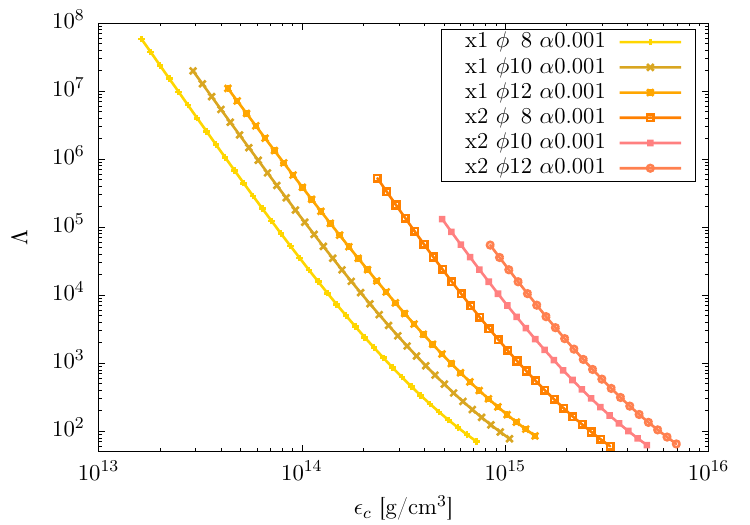}
    ~\caption{Tidal deformability $\Lambda$ vs. central energy density $\epsilon_c$ for the fermionic dark matter EoSs
    considered in this study. The EoSs are labelled according to the value of their fermion mass $m_{\textrm{X}}=(1,2)$ GeV
    and the mass of the mediator $m_{\phi}=(8,10,12)$ MeV. The fine structure constant is fixed at $\alpha=10^{-3}$.
    For each curve, the lowest value of $\Lambda$ corresponds to the tidal deformability of the last stable model, while
the highest value identifies a stellar model with mass equal to $M=0.5\,M_{\odot}$.}
    \label{fig:fdm-lambda-vs-eps}
\end{figure}

Similarly, Fig.~\ref{fig:fdm-lambda-vs-eps} shows the behaviour of the tidal deformability $\Lambda$ as a function of the central energy
density $\epsilon_c$ for some fermionic dark matter EoSs.
For each curve, the lowest value of $\Lambda$ corresponds to the tidal deformability of the last stable model, while
the highest value identifies a stellar model with mass equal to $M=0.5\,M_{\odot}$.
We observe that the value of $\Lambda$ increases when the mediator mass $m_{\phi}$ is increased.
A similar behaviour for $\Lambda$ is obtained by increasing the mass of the dark fermion $m_{\textrm{X}}$.
Given the results shown in Fig.~\ref{fig:fdm-lambda-vs-eps}, we choose to investigate the
validity of Chan and Kuan's relations (see Refs.~\cite{2014PhRvD..90l4023C, 2022MNRAS.513.4045K}) in a range of
values of $\Lambda \in [100, 12000]$, which provides stable stellar models
heavier than $M=0.5\,M_{\odot}$.

\begin{figure}
    \centering
    \includegraphics[width=0.5\textwidth]{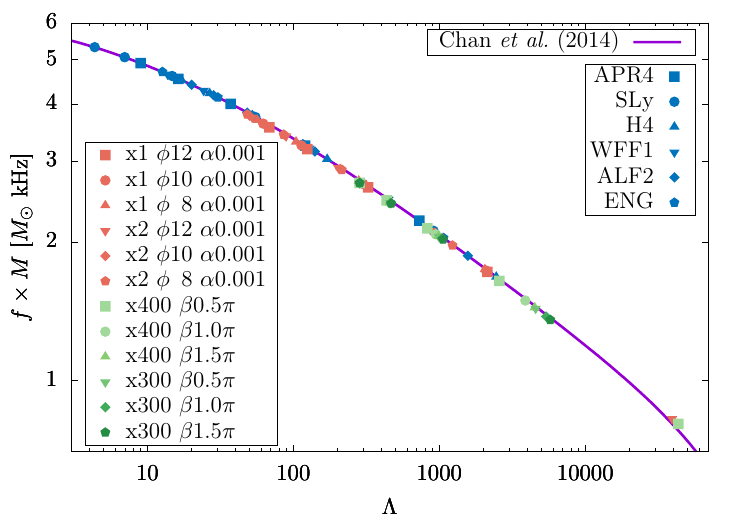}
    ~\caption{The mass-scaled $f$-mode frequency is plotted against the tidal deformability $\Lambda$.
        The curve represents the universal relation established by Chan $\etal$~\cite{2014PhRvD..90l4023C},
        and reported in Eq.~\eqref{eq:chan}.
    Green, orange and blue points represent stellar models as described in
    Fig.~\ref{fig_f_AK}.}
    \label{fig:chan}
\end{figure}

\begin{figure}
    \centering
    \includegraphics[width=0.5\textwidth]{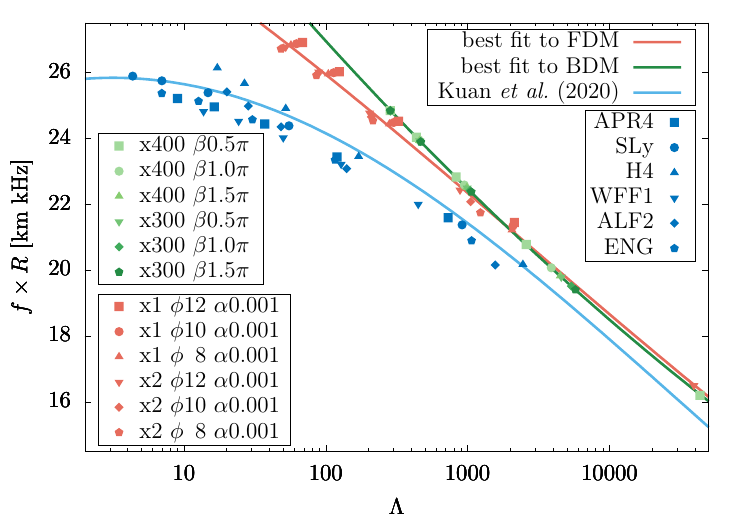}
    ~\caption{The radius-scaled $f$-mode frequency is plotted against the tidal deformability $\Lambda$.
    The curve represents the universal relation established in Kuan $\etal$~\cite{2022MNRAS.513.4045K}
    and reported in Eq.~\eqref{eq:kuan}.
    The universal relation for fermionic dark matter found in this paper is shown as an orange curve, which corresponds
    to Eq.~\eqref{eq:kuan-fdm}. Bosonic dark matter is best represented by Eq.~\eqref{eq:kuan-bdm}, which is displayed
    as a green curve.
    Green, orange and blue points represent stellar models as described in
    Fig.~\ref{fig_f_AK}.}
    \label{fig:kuan}
\end{figure}

The universal relation proposed by Chan $\etal$~\cite{2014PhRvD..90l4023C} establishes a connection
between the mass-scaled $f$-mode frequency of NSs and their tidal deformability $\Lambda$, and reads\footnote{Here and in the following, we denote with
``$\log$'' the decadic logarithm.}
\begin{equation}
    \log\left(\frac{f}{\textrm{kHz}} \frac{M}{M_\odot}\right) = 0.814 -0.050(\log(\Lambda)) - 0.035(\log(\Lambda))^2.
    \label{eq:chan}
\end{equation}
The relation above is shown in Fig.~\ref{fig:chan} with a purple curve.
We observe that the given relation applies not only to NS models, but also fits the behavior of dark bosonic and
fermionic stars nicely.

Additionally, another relation between the radius-scaled $f$-mode frequency and $\Lambda$ was found by Kuan $\etal$~\cite{2022MNRAS.513.4045K}, and this is given by
\begin{equation}
    \log\left(\frac{f}{\textrm{kHz}} \frac{R}{\textrm{km}}\right) = 1.409 + 0.013(\log(\Lambda)) -0.013 (\log(\Lambda))^2.
    \label{eq:kuan}
\end{equation}

Kuan's relation is shown in Fig.~\ref{fig:kuan} as a light-blue curve.
It can be seen that fermionic and bosonic dark matter models do not follow Eq.~\eqref{eq:kuan}.
For this reason, we propose other quadratic fits for the radius-scaled $f$-mode frequency.
The best fit to fermionic dark matter data reads
\begin{equation}
    \log\left(\frac{f}{\textrm{kHz}} \frac{R}{\textrm{km}}\right) = 1.503 - 0.0311
    (\log(\Lambda)) -0.00672(\log(\Lambda))^2.
    \label{eq:kuan-fdm}
\end{equation}
The relative error of Eq.~\eqref{eq:kuan-fdm} with respect to the stellar
models considered is smaller than 1$\%$.

Bosonic dark matter data are best fitted by the following relation instead
\begin{equation}
    \log\left(\frac{f}{\textrm{kHz}} \frac{R}{\textrm{km}}\right) = 1.572 - 0.0652 (\log(\Lambda)) - 0.00275(\log(\Lambda))^2,
    \label{eq:kuan-bdm}
\end{equation}
which reproduces the behaviour of the given data with a relative error smaller
than 1$\%$.

\section{Inversion scheme}
\label{sec:inversion-scheme}
Assuming that both the oscillation frequency $f$ and the damping time $\tau$ of the $f$-mode of a dark star have been obtained from
GW observations, we can combine the universal relations outlined in this paper to estimate some macroscopic properties
of the star.
For example, by solving Eq.~\eqref{UR_Lau_freq} together with Eq.~\eqref{eq_UR_damp_myLLL_bdm}, we might estimate
the mass $M$ and the moment of inertia $I$ of a bosonic dark matter star.
Analogously, by combining Eq.~\eqref{UR_Lau_freq} and Eq.~\eqref{eq_UR_damp_myLLL_fdm}, an estimate of the
mass $M$ and the moment of inertia $I$ of a fermionic dark matter star can be obtained.
Let us write Eq.~\eqref{UR_Lau_freq} in the general form
\begin{equation}
    M \omega_r = a_1 + a_2 \eta + a_3 \eta^2,
\end{equation}
where $a_1=-0.0047$, $a_2=0.133$ and $a_3=0.575$.
We then isolate $M$ on the left-hand side of the equation, so that we get
\begin{equation}
    M = \frac{a_1}{\omega_r} + \frac{a_2}{\omega_r} \eta + \frac{a_3}{\omega_r} \eta^2.
    \label{eq:lau-inverse-freq}
\end{equation}
Moreover, since Eq.~\eqref{eq_UR_damp_myLLL_bdm} and~\eqref{eq_UR_damp_myLLL_fdm} have the same structure and differ
only in the values of the coefficients, let us consider the general form
\begin{equation}
    \frac{I^2}{M^5} \omega_i = b_1 + b_2\eta^2
    \label{eq:lau_general}
\end{equation}
where $b_1$ and $b_2$ are coefficients that depend on the specific type of matter considered, and whose values are summarized
in Table~\ref{table:coefficients_LLL_inverse_general}.
\begin{table*}
    \centering
    \caption{Coefficients to be used in the proposed inversion scheme based on the universal relations of
    Ref.~\cite{Lau_2010} in case of neutron matter, bosonic or fermioni dark matter.}
    \ra{1.3}
    \begin{tabular}{@{}ccc@{}}
        \toprule
        EoS & $b_1$ & $b_2$ \\
        \hline
        NS & 0.00694 & -0.0256 \\
        BDM & 0.006858 & -0.03105 \\
        FDM & 0.006604 & -0.02521\\
        \bottomrule
    \end{tabular}
    \label{table:coefficients_LLL_inverse_general}
\end{table*}

We isolate the mass on the left-hand side, rewriting Eq.~\eqref{eq:lau_general} as
\begin{equation}
    M = \frac{b_1}{\omega_i} \eta^4 + \frac{b_2}{\omega_i} \eta^6.
    \label{eq:lau-inverse-damp}
\end{equation}

\begin{figure}
    \centering
    \includegraphics[width=0.5\textwidth]{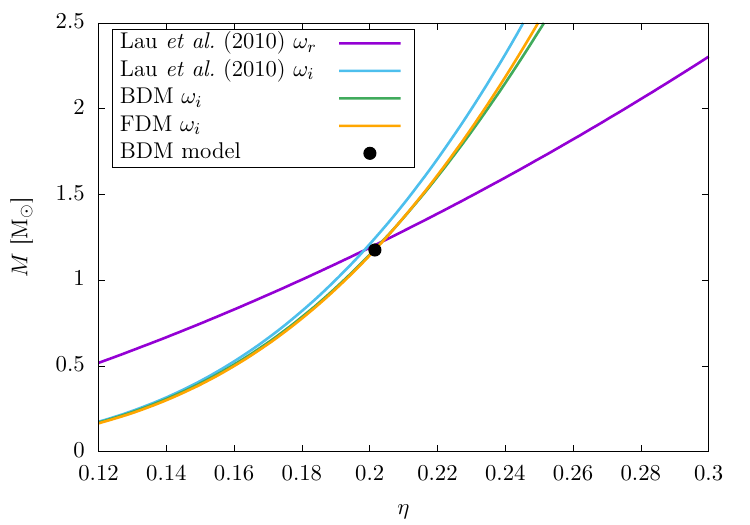}
    ~\caption{An illustration of how accurately the moment of inertia and the mass of a bosonic dark matter star can be inferred
    from detected $f$-mode data and universal relations of Eqs.~\eqref{UR_Lau_freq} (purple curve) and~\eqref{eq_UR_damp_myLLL_bdm}
    (green curve). The true values of the BDM model considered are represented with a black point.
    The universal relation for the damping time of FDM stars of Eq.~\eqref{eq_UR_damp_myLLL_fdm} is also included (orange
        curve) for comparison, as well as Eq.~\eqref{UR_Lau_damp} (light-blue
        curve) for the damping time of NS models.}
    \label{fig:inversion-LLL}
\end{figure}

By substituting the value of $M$ from Eq.~\eqref{eq:lau-inverse-freq} on the left-hand side of
Eq.~\eqref{eq:lau-inverse-damp}, we get an
estimate of the effective compactness $\eta = \sqrt{M^3/I}$, which can then be plugged back into Eq.~\eqref{eq:lau-inverse-freq}
to evaluate the mass $M$ and the moment of inertia $I$ of the source of the measured $f$-mode frequency.

A graphical solution to the system of Eqs.~\eqref{eq:lau-inverse-freq} and~\eqref{eq:lau-inverse-damp} is shown
in Fig.~\ref{fig:inversion-LLL} for the case of a bosonic dark matter model,
where the two equations are represented with a purple and a green curve, respectively.
One should note that these curves represent an ideal case,
    while the lines would be thicker if uncertainties were taken into account.
The intersection point of the two curves corresponds to the estimates of the mass $M$ and moment of inertia $I$ of a bosonic source
that can be achieved by using the universal relations~\eqref{eq:lau-inverse-freq} and~\eqref{eq:lau-inverse-damp},
with values for the coefficients $a_1$, $a_2$, $a_3$, $b_1$, $b_2$ given by those of Eq.~\eqref{UR_Lau_freq} and~\eqref{eq_UR_damp_myLLL_bdm}.
In order to compare the estimates with the true values of the bosonic dark matter model of which we assume to have measured
the $f$-mode frequency and damping time, we plot the model's mass versus effective compactness $\eta = \sqrt{M^3/I}$
as a black point.
It can be seen that the inversion scheme based on the universal relations provides a very good estimate of the stellar
parameters.

In principle, assuming to have measured the $f$-mode frequency and damping time of a source, we would like to be able to
infer the type of matter it is composed of, namely to distinguish between neuron matter, bosonic and fermionic dark matter.
In order to explore the feasibility of the idea, we add two more curves to Fig.~\ref{fig:inversion-LLL}, corresponding
to the universal relation for the $f$-mode damping time of neutron matter of
Ref.~\cite{Lau_2010} (cf. Eq.~\eqref{UR_Lau_damp}) in light-blue 
and to our newly determined Eq.~\eqref{eq_UR_damp_myLLL_fdm} for the damping
time of fermionic dark matter in orange, respectively.
We can see from Fig.~\ref{fig:inversion-LLL} that extremely accurate
measurements of the $f$-mode are necessary in order to distinguish between the
different types of matter.

\begin{figure}
    \centering
    \includegraphics[width=0.5\textwidth]{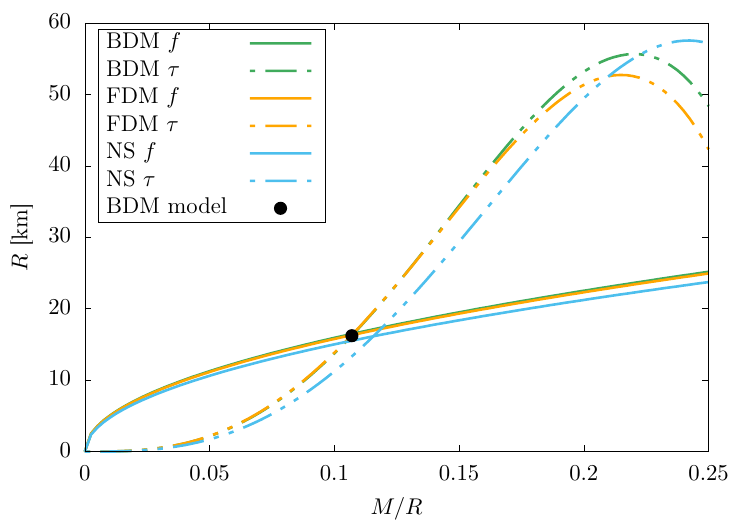}
    ~\caption{An illustration of how accurately the radius and the mass of a bosonic dark matter star can be inferred
    from detected $f$-mode data and universal relations of Eq.~\eqref{UR_bdm_myAKstyle} (green curve)
    and~\eqref{UR_bdm_damping_myAKstyle} (green dashed-dotted curve).
    The true values of the BDM model considered are represented with a black point.
    For comparison, we also include the universal relations for FDM given by Eq.~\eqref{UR_fdm_myAKstyle} (orange curve)
    and~\eqref{UR_fdm_damping_myAKstyle} (orange dashed-dotted curve), as well as those for NS matter given by Eq.~\eqref{UR_AK_frequency_my} (light-blue curve)
    and~\eqref{UR_AK_damping_my} (light-blue dashed-dotted curve).}
    \label{fig:inversion-AK}
\end{figure}

So far, we have only considered the universal relations provided by Lau $\etal$ in Ref.~\cite{Lau_2010} and similar ones
obtained by finding the best linear fits to bosonic and fermionic dark matter models.
However, the universal relations by Andersson and Kokkotas~\cite{1998MNRAS.299.1059A} can also be used
to estimate some macroscopic properties of a compact star.
Indeed, assuming to have measured the $f$-mode frequency and damping time of oscillations of a given neutron star,
we might combine Eqs.~\eqref{UR_AK_frequency_my} and~\eqref{UR_AK_damping_my} to estimate the radius $R$ and the compactness
$M/R$ of the source and, consequently, also its mass $M$.
If the source is a bosonic DM model instead, we could apply the same
procedure, but considering Eqs.~\eqref{UR_bdm_myAKstyle}
and~\eqref{UR_bdm_damping_myAKstyle}.
Finally, if the source is a fermionic DM model, the inversion scheme should be applied
to Eqs.~\eqref{UR_fdm_myAKstyle} and~\eqref{UR_fdm_damping_myAKstyle}.
Since the inversion scheme is the same for all types of matter considered (i.e., neutron matter, bosonic dark matter, or fermionic dark matter), differing only in the coefficients of the universal relations that depend on the matter type, we present below a possible inversion procedure using general forms for the equations.
The specific values assumed by the coefficients in case of neutron, bosonic and fermionic dark matter are then reported
in Table~\ref{table:coefficients_AK_inverse_general}.
The system of equations to be solved is given by the general form of the
Andersson and Kokkotas universal relations
of Eqs.~\eqref{UR_AK_frequency_my} and~\eqref{UR_AK_damping_my}, namely
\begin{equation}
    \begin{cases}
        f = c_1 + c_2 \sqrt{\frac{M}{R^3}}, \\
        \frac{R^4}{M^3\tau} = d_1 + d_2 \frac{M}{R}.
    \end{cases}
    \label{eq:AK_inverse_general}
\end{equation}
\begin{table*}
    \centering
    \caption{Coefficients to be used in the proposed inversion scheme based on the universal relations of
    Ref.~\cite{1998MNRAS.299.1059A} in case of neutron matter, bosonic or
    fermionic dark matter.}
    \ra{1.3}
    \begin{tabular}{@{}ccccc@{}}
        \toprule
        EoS & $c_1$ & $c_2$ & $d_1$ & $d_2$ \\
        \hline
        NS & 0.22 & 32.16 & 0.08627 & -0.2676 \\
        BDM & -0.01461 & 42.11 & 0.1105 & -0.3764 \\
        FDM & 0.06661 & 38.97 & 0.1126 & -0.3930\\
        \bottomrule
    \end{tabular}
    \label{table:coefficients_AK_inverse_general}
\end{table*}
By taking $R$ out of the square root on the right-hand side of the first equation, and isolating $R$ on the left hand
side of the second, we get
\begin{equation}
    \begin{cases}
        f = c_1 + c_2 \frac{1}{R}\sqrt{\frac{M}{R}}, \\
        R \left(\frac{R}{M}\right)^3\frac{1}{\tau} = d_1 + d_2 \frac{M}{R}.
    \end{cases}
    \label{eq:AK_inverse_general_intermediate_step}
\end{equation}
Finally, by reshuffling the terms, we can express $R$ in terms of the stellar compactness $M/R$ in the following way
\begin{equation}
    \begin{cases}
        R = \frac{c_2}{f - c_1} \sqrt{\frac{M}{R}}, \\
        R = d_1 \tau \left( \frac{M}{R}\right)^3 + d_2 \tau \left( \frac{M}{R}\right)^4.
    \end{cases}
    \label{eq:AK_inverse_general_final}
\end{equation}
The aforementioned system of Eq.~\eqref{eq:AK_inverse_general_final} can be solved graphically.
An example of that is shown in Fig.~\ref{fig:inversion-AK} for the case of a bosonic DM model.
The true values of the radius $R$ and compactness $M/R$ of the chosen model are displayed with a black point.
By plotting the system of Eq.~\eqref{eq:AK_inverse_general_final} with the correct coefficients for BDM, we find
the green solid and dashed-dotted curves for the first and the second equation, respectively.
Similar to the case of the inversion scheme based on Lau~\etal-like universal relations,
 these curves show an ideal case, but the lines would be thicker if uncertainties were considered.
It can be stated that the proposed inversion scheme works very well for estimating the stellar macroscopic parameters
involved.
For comparison, we also include, in the same plot, the curves that we would get by considering the same measured values
of the $f$-mode frequency and damping time, but assuming that the source is made of neutron or fermionic dark matter.
In particular, the orange solid and dashed-dotted curves refer to FDM, while the light-blue solid and dashed-dotted
ones to NS matter.
From the results of this plot (Fig.~\ref{fig:inversion-AK}) we observe that an $f$-mode observation of
decent accuracy should allow to distinguish between nuclear and dark matter.
In the case of dark matter, these equations, however, do not allow to
distinguish between fermionic and bosonic matter in particular; other
observations or relations would be necessary to decide this.

\begin{table*}
    \centering
    \caption{The first three columns show mass $M$, radius $R$, and moment of inertia $I$ of the selected models;,
    columns five and six show
    their $f$-mode frequencies $f$ and damping times $\tau$, respectively.
    The last four columns show
    relative errors of estimated quantities obtained from the $f$-mode data by making use of
    universal relations. Columns seven and eight (labeled AK98) and nine and
    ten (labeled LLL10) are based on the universal relations
    by Andersson and Kokkotas \cite{1998MNRAS.299.1059A} and
    Lau~\etal~\cite{Lau_2010}, respectively.}
    \ra{1.3}
    \begin{tabular}{@{}cccccc|cc|cc@{}}
        \toprule
        EoS & $M$ & $R$ & $I$ & $f$ & $\tau$ & $\delta M/M$ & $\delta R/R$ &
        $\delta I/I$ & $\delta M/M$\\
         & $[M_\odot]$ & [km] & [$10^{45}\,\textrm{g}\,\textrm{cm}^2$] & [kHz] & [ms] & \multicolumn{2}{c|}{AK98} & \multicolumn{2}{c}{LLL10} \\
        \hline
        $x300 \beta0.5\pi$      & 1.176 & 16.23 & 1.736 & 1.220 &  630 & 0.026 & 0.019 & 0.108 & 0.059 \\
        $x300 \beta1.0\pi$      & 1.634 & 23.15 & 4.906 & 0.843 &  958 & 0.018 & 0.013 & 0.103 & 0.040 \\
        $x300 \beta1.5\pi$      & 1.987 & 28.46 & 9.007 & 0.682 & 1209 & 0.012 & 0.009 & 0.102 & 0.039 \\
        $x400 \beta0.5\pi$      & 0.500 & 10.07 & 0.278 & 1.609 & 1027 & 0.052 & 0.037 & 0.047 & 0.021 \\
        $x400 \beta1.0\pi$      & 0.951 & 12.79 & 0.873 & 1.570 &  466 & 0.029 & 0.021 & 0.111 & 0.043 \\
        $x400 \beta1.5\pi$      & 1.303 & 14.23 & 1.497 & 1.582 &  322 & 0.014 & 0.009 & 0.164 & 0.065 \\
        $x1 \phi\phantom{0}8 \alpha0.001$  & 2.270 & 27.62 & 9.803 & 0.768 &  789 & 0.067 & 0.041 & 0.109 & 0.046 \\
        $x1 \phi10 \alpha0.001$            & 1.878 & 23.09 & 5.579 & 0.924 &  661 & 0.032 & 0.007 & 0.090 & 0.037 \\
        $x2 \phi\phantom{0}8 \alpha0.001$  & 1.350 & 10.19 & 0.899 & 2.542 &  129 & 0.010 & 0.001 & 0.112 & 0.039 \\
        \bottomrule
    \end{tabular}
    \label{table:inversion-scheme}
\end{table*}

Since all the universal relations considered hold only approximately, the inferred values of $M$, $R$ and $I$ are
expected to deviate from the exact values by $\delta M$, $\delta R$ and
$\delta I$, respectively.
To validate the inversion schemes outlined above and test their accuracy, we assume to have measured the $f$-mode
frequency $f$ and damping time $\tau$ of several bosonic and fermionic dark matter stellar models and we try to numerically estimate
their mass $M$, radius $R$ and moment of inertia $I$ by making use of the universal relations associated to the specific
type of matter.
Table~\ref{table:inversion-scheme} shows the EoS defining the stellar models, the true values of their mass $M$, radius
$R$, as well as $f$-mode frequency $f$ and damping time $\tau$; further, the
relative errors of the estimated mass, radius and moment of inertia obtained
from the universal relations are listed in the last four columns.
More precisely, columns six and seven refer to results of the inversion
scheme based on universal relations of Andersson and Kokkotas~\cite{1998MNRAS.299.1059A}, while the last two
columns refer to the scheme based on the universal relations
from Lau $\etal$~\cite{Lau_2010}.
In general, the inversion scheme works very well for both bosonic and fermion dark matter EoSs.
The relative error in the estimates of mass and radius are of the order of few percent or better, while the estimate of
the moment of inertia is less accurate, with a relative error of the order of 10\%.

\section{Conclusions}
\label{sec:conclusions}
In this paper we have investigated compact stars modeled with fermionic and bosonic DM EoSs, as viable candidates to be
tested with future GW observations.
By solving the stellar structure for static, spherically-symmetric bodies, we have derived some basic bulk properties
of such dark stars, namely their mass and radius.
We have compared these results with some well known nuclear EoSs, showing that dark objects may cover large portions
of the parameter space close to standard NSs.

We have also tested the validity of some universal relations connecting the $f$-mode oscillation frequency and
damping time to the mass, radius, moment of inertia and tidal deformability of compact objects.
These relations hold true for a large variety of NS EoSs, but in most cases
dark stellar models deviate to some extent from them.
Consequently, we propose alternative forms of those relations, which best fit the dark sector.
More precisely, we establish a pair of empirical equations, namely Eqs.~\eqref{UR_bdm_myAKstyle} and~\eqref{UR_bdm_damping_myAKstyle},
which can predict the frequency and the damping time of $f$-modes from $M$ and $R$ with good accuracy for BDM.
An analogous approach is followed for determining those quantities with FDM,
leading to Eqs.~\eqref{UR_fdm_myAKstyle}
and~\eqref{UR_fdm_damping_myAKstyle}.
Additionally, we study the universality embedded in the moment of inertia of dark objects, finding 
Eq.~\eqref{eq_UR_damp_myLLL_bdm} for bosonic dark matter, and the corresponding Eq.~\eqref{eq_UR_damp_myLLL_fdm}
for fermionic dark matter.
The universality encoded by the aforementioned relations leads to an accurate inversion scheme.

Finally, we take into account the tidal deformability characterizing the dark objects if they existed in binary systems,
and we explore the validity of some universal relations previously established for NSs.
We find that the universal relation for the mass-scaled $f$-mode frequency
holds for all types of stars, while the radius-scaled $f$-mode frequency
acquires different universal relations for the different types of matter;
we encode those in Eqs.~\eqref{eq:kuan-bdm}
and~\eqref{eq:kuan-fdm} for bosonic and fermionic dark matter, respectively.

As an application, we present a simple inversion scheme to demonstrate
how the competing universal relations for the different types of matter can be
inverted, given some potential observations; distinguishing between
different types of matter is, in principle, possible, however, sufficiently
accurate observations are necessary. A more thorough exploration of the
parameter space of dark stars could be conducted by employing Bayesian
inference which, however, is beyond the scope of the present paper and we keep
this extension for future work.

\begin{acknowledgments}
This project has received funding from the European Union’s Horizon-MSCA-2022 research and innovation programme ``Einstein Waves’’ under grant agreement No. 101131233.
\end{acknowledgments}

\section*{Data availability}
The data that support the findings of this article are not publicly available because no suitable repository exists for hosting data in this field of study. The data are available from the authors upon reasonable request.

\appendix
\section{Equilibrium stellar model near $r=R$}
\label{taylor-expansion}
To calculate the pulsation of a stellar model, one must first numerically determine the equilibrium stellar model.
The general static spherical metric, which describes the geometry of an equilibrium stellar model,
can be expressed as follows:
\begin{equation}
    \dif s^2 = - e^{\nu} \dif t^2 + e^{\lambda} \dif r^2 + r^2(\dif\theta^2 +
    \sin^2\theta \dif\phi^2 ),
\end{equation}
where the metric potentials $\nu$ and $\lambda$ are functions of $r$.
It might be useful to define the function $M(r)$ as follows:
\begin{equation}
    M(r) = \frac{1}{2}r(1-e^{-\lambda}).
\end{equation}
Given this metric, Einstein's equations correspond to the very well known Tolman-Oppenheimer-Volkoff equations, namely
\begin{align}
    \frac{dM}{dr} &= 4 \pi r^2 \epsilon \nonumber\\
    \frac{d\nu}{dr} &= 2 \frac{M + 4\pi r^3 P}{r(r-2M)} \\
    \frac{dP}{dr} &= - \frac{1}{2}(\epsilon + P) \frac{d\nu}{dr}. \nonumber
    \label{eq:TOV}
\end{align}
where $\epsilon$ and $P$ are the energy density and the pressure of the fluid in the star, respectively.
The pressure and the density are related by an equation of state $\epsilon = \epsilon(P)$.
These equations are numerically integrated as an initial value problem starting from the center of the star, $r=0$.
At the center, the mass function must be zero, $M(0)=0$, while the values of $P$ and $\nu$ can be freely specified.
The integration continues up until the radius $r$ at which the pressure vanishes, $P(r)=0$.
We denote this radius by $R$ and it represents the surface of the star.
The total mass of the star is given by $M(R)$, and the arbitrary constant in the function $\nu$ is determined by normalizing
the time coordinate at spatial infinity:
\begin{equation}
    e^{\nu(R)} = 1- \frac{2M(R)}{R}.
\end{equation}

Approaching the star's surface, pressure values approach zero.
Considering the fermionic dark matter EoS, cf. Eqs.~\eqref{eps_eos_fermions} and~\eqref{p_eos_fermions}, we point out that,
for small values of $P$ and, consequently, of $x$, the two relations suffer from numerical cancellation.
Consequently, a power series expansion of both $\epsilon(x)$ and $P(x)$ around $x=0$ is required.
More precisely, the relevant quantities to be Taylor expanded are the functions $\xi(x)$ and $\chi(x)$, which are contained
in the definition of $\epsilon(x)$ and $P(x)$.
Starting from $\xi(x)$, the following Taylor expansion should be considered
for $x \lesssim 0.19$:
\begin{align}
    \xi(x) \approx & \frac{8}{3} x^3 + \frac{4}{5} x^5 -\frac{1}{7} x^7 \nonumber\\
    &+ \frac{1}{18} x^9 - \frac{5}{176} x^{11} +\frac{7}{416} x^{13} \nonumber\\
    &- \frac{7}{640} x^{15} + \frac{33}{4352} x^{17} - \frac{429}{77824} x^{19}.
\end{align}
The threshold $x\simeq 0.19$ has been computed equating the smallest contribution given by the Taylor expansion to
machine precision, assuming the latter to be $\textrm{EPS} \approx 10^{-16}$, namely
\begin{equation}
    \textrm{EPS} \approx \frac{429}{77824}x^{19}.
\end{equation}
As regards $\chi(x)$, we find that the following power series expansion holds
for $x \lesssim 0.27$:
\begin{align}
    \chi(x) \approx & \frac{8}{15} x^5 - \frac{4}{21} x^7 + \frac{1}{9} x^9 \nonumber\\
    &- \frac{5}{66} x^{11} + \frac{35}{624} x^{13} - \frac{7}{160} x^{15} \nonumber \\
    &+ \frac{77}{2176} x^{17} - \frac{143}{4864} x^{19} + \frac{715}{28672} x^{21} \nonumber \\
    &- \frac{12155}{565248} x^{23} + \frac{46189}{2457600} x^{25}.
\end{align}
Similarly to what is done for $\xi(x)$, also in this case the threshold for $x$ has been computed relating the magnitude of the
smallest contribution of the series to machine precision.

Let us now consider the EoS for bosonic dark matter given by Eq.~\eqref{eos_bdm}.
By defining the quantity $\epsilon_0 \equiv m_{\textrm{X}}^4 c^3 / (4 \pi \beta \hbar^3)$ and setting $x \equiv \epsilon / \epsilon_0$,
the EoS can be rewritten as
\begin{equation}
    P = \frac{1}{9} \epsilon_0 \left( \sqrt{4+3x} -2 \right)^2.
\end{equation}
Let us assign the content of the parenthesis to an auxiliary variable
$F\equiv\sqrt{4+3x} -2$. At low energy densities, the calculation of $F$ is numerically unstable.
Consequently, for small $x$, a Taylor expansion should be used. While the
numerical cancellation in the expression
for the pressure can be avoided by multiplying out $F^2$, the variable $F$
itself appears in other expressions, making a Taylor expansion indeed useful.
More precisely, we find that, for $x \lesssim 10^{-4}$, it holds
\begin{align}
    F \approx & \frac{3}{4} x - \frac{9}{64} x^2 + \frac{27}{512} x^3 \nonumber\\
    &- \frac{405}{16384} x^4 + \frac{1701}{131072} x^5.
\end{align}

The presented Taylor expansions allow not only a more accurate determination
of the stellar models in the crustal part, but due to the higher accuracy that
is achieved, the Runge-Kutta integrator is allowed to make larger steps which
considerably speeds up the calculation.

\section{Self-similarity of the bosonic DM EoS}
\label{sec:self-similarity}
In this appendix we present the features of the EoS which lead to
self-similarity of the bosonic dark matter EoS, i.e. the shape of the mass-radius
relation is independent of EoS parameters such as the mass of the dark particle $m_\textrm{X}$ and the coupling constant
$\beta$.

The following derivation is closely based on, and some parts are directly reproduced from,
the mathematical proof given in Ref.~\cite{2017PhRvD..96b3005M}.
To begin, we define the following quantities corresponding to a dimensionless mass and radius:
\begin{equation}
    M_* = \frac{GM}{c^2l}
	\quad\text{and}\quad
	r_* = \frac{r}{l},
\end{equation}
where $l$ is a length scale. 
Moreover, we use this scaling $l$ to define a dimensionless density $\epsilon_*$ and pressure $P_*$
\begin{equation}
    \epsilon_* = \frac{l^2G}{c^2}\epsilon
	\quad\text{and}\quad
	P_* = \frac{l^2G}{c^4}P.
\end{equation}
Using these variables, the TOV equations can be cast in the following dimensionless form
\begin{align}
    \frac{dP_*}{dr_*} &= - \frac{M_* \epsilon_*}{r^2_*}\left( 1 +
    \frac{P_*}{\epsilon_*} \right) \left( 1 + \frac{4\pi r_*^3 P_*}{M_*} \right)
    \left( 1- \frac{2M_*}{r_*}\right)^{-1}, \nonumber\\
    \frac{dM_*}{dr_*} &= 4\pi r_*^2 \epsilon_*.
\end{align}

Given an EoS, the mass-radius relation follows by varying the central energy
density $\epsilon_*(0)$.
Parameters of the model such as the dark particle mass $m_\textrm{X}$ and the coupling constant $\beta$ can affect the TOV equations
only if they are contained in the EoS.
However, it is possible to choose the scaling parameter $l$ such that the EoS written in dimensionless variables is
independent of the model parameters.
This happens when the scaling factor is defined as
\begin{equation}
    l = \sqrt{\frac{3\hbar ^3 \beta}{Gc}} \frac{1}{m_\textrm{X}^2},
\end{equation}
and the EoS is given by
\begin{equation}
    P_* = \frac{1}{3}(\sqrt {1+\epsilon_*} - 1)^2.
\end{equation}

We have repeated the self-similarity of the bosonic DM EoS here because it
provides a simple argument that we expect certain universal relations based on
bosonic EoS to be linear functions. These are the relations concerning the
$f$-mode frequency or damping time as a function of compactness or effective
compactness.

\section{Results for various equations of state}
\label{sec:fake-appendix}
This appendix provides the numerical data for mode frequencies of all the EoSs considered in the present study.
These data were used to infer the empirical relations discussed in the main body of the paper.
We provide the data in the form of one table for each type of matter, namely neutron matter (c.f., Table~\ref{tab:ns}), bosonic dark matter (c.f., Table~\ref{tab:table-bdm}) and fermionic
dark matter (c.f., Table~\ref{tab:table}).
In each table we list the EoS, the central energy density, the mass and the
radius of the stellar model, the frequency
and damping time of the $f$-mode, the moment of inertia and the tidal deformability of the compact object.

\onecolumngrid
\begin{table*}[h]
    \centering
    \caption{Data for EoSs of neutron matter. The columns are the identifier
    of the EoS, the energy density $\epsilon_c$, the mass $M$, the radius $R$,
    the $f$-mode's frequency $f$ and damping time $\tau$, the stars's moment
    of inertia $I$ and its tidal deformability $\Lambda$, respectively.}
    \ra{1.3}
    \begin{tabular}{@{}cccccccc@{}}
        \toprule
        EoS & $\epsilon_c$ & $M$ & $R$ & $f$ & $\tau$ & $I$ & $\Lambda$ \\
        \hline
        --- & [$10^{15}$g$/$cm$^3]$ & $[M_{\odot}]$ & [km] & [kHz] & [ms] &
        [$10^{45}$g\,cm$^2$] & --- \\
        \hline
        \multirow{4}{4em}{APR4}
        & 0.882 & 1.170 & 11.34 & 1.903 & 235.9 & 1.010 & 7.28e+02 \\
        & 1.112 & 1.561 & 11.27 & 2.078 & 155.4 & 1.523 & 1.20e+02 \\
        & 1.341 & 1.819 & 11.11 & 2.199 & 134.3 & 1.878 & 3.70e+01 \\
        & 1.571 & 1.980 & 10.90 & 2.289 & 131.0 & 2.089 & 1.63e+01 \\
        & 1.801 & 2.078 & 10.68 & 2.359 & 135.0 & 2.196 & 9.01e+00 \\
        \hline
        \multirow{4}{4em}{SLy}
        & 0.833 & 1.170 & 11.81 & 1.809 & 262.0 & 1.074 & 9.13e+02 \\
        & 1.340 & 1.745 & 11.36 & 2.145 & 140.0 & 1.792 & 5.49e+01 \\
        & 1.847 & 1.960 & 10.82 & 2.345 & 128.8 & 1.989 & 1.47e+01 \\
        & 2.354 & 2.033 & 10.35 & 2.486 & 133.5 & 1.977 & 6.99e+00 \\
        & 2.861 & 2.048 & 9.97  & 2.595 & 141.4 & 1.899 & 4.35e+00 \\
        \hline
        \multirow{4}{4em}{H4}
        & 5.209 & 1.170 & 14.05 & 1.435 & 430.1 & 1.428 & 2.45e+03 \\
        & 0.933 & 1.742 & 13.47 & 1.739 & 195.0 & 2.292 & 1.70e+02 \\
        & 1.345 & 1.936 & 12.76 & 1.950 & 154.2 & 2.415 & 5.22e+01 \\
        & 1.758 & 2.000 & 12.18 & 2.107 & 140.9 & 2.331 & 2.67e+01 \\
        & 2.171 & 2.013 & 11.69 & 2.234 & 135.5 & 2.196 & 1.71e+01 \\
        \hline
        \multirow{4}{4em}{WFF1}
        & 1.061 & 1.170 & 10.38 & 2.118 & 190.3 & 0.887 & 4.48e+02 \\
        & 1.230 & 1.436 & 10.41 & 2.230 & 146.0 & 1.206 & 1.28e+02 \\
        & 1.398 & 1.642 & 10.37 & 2.316 & 129.0 & 1.468 & 5.01e+01 \\
        & 1.567 & 1.793 & 10.28 & 2.384 & 123.7 & 1.665 & 2.42e+01 \\
        & 1.736 & 1.903 & 10.17 & 2.439 & 124.6 & 1.803 & 1.37e+01 \\
        \hline
        \multirow{4}{4em}{ALF2}
        & 0.603 & 1.170 & 12.62 & 1.596 & 341.0 & 1.254 & 1.56e+03 \\
        & 0.986 & 1.711 & 12.53 & 1.841 & 178.8 & 2.077 & 1.40e+02 \\
        & 1.369 & 1.906 & 12.12 & 2.008 & 148.4 & 2.273 & 4.83e+01 \\
        & 1.753 & 1.963 & 11.74 & 2.126 & 138.6 & 2.238 & 2.84e+01 \\
        & 2.136 & 1.976 & 11.40 & 2.227 & 133.6 & 2.139 & 2.00e+01 \\
        \hline
        \multirow{4}{4em}{ENG}
        & 0.745 & 1.170 & 11.96 & 1.746 & 281.9 & 1.122 & 1.06e+03 \\
        & 1.028 & 1.663 & 11.88 & 1.964 & 163.7 & 1.827 & 1.16e+02 \\
        & 1.312 & 1.952 & 11.60 & 2.116 & 139.1 & 2.235 & 3.03e+01 \\
        & 1.596 & 2.108 & 11.28 & 2.226 & 137.4 & 2.417 & 1.26e+01 \\
        & 1.880 & 2.187 & 10.97 & 2.311 & 143.5 & 2.467 & 6.97e+00 \\
        \bottomrule
    \end{tabular}
    \label{tab:ns}
\end{table*}

\begin{table*}[h]
    \centering
    \caption{Data for EoSs of bosonic dark matter. The columns are the same as
    in Tab.~\ref{tab:ns}.}
    \ra{1.3}
    \begin{tabular}{cccccccc}
        \toprule
        EoS & $\epsilon_c$ & $M$ & $R$ & $f$ & $\tau$ & $I$ & $\Lambda$ \\
        \hline
        --- & [$10^{15}$g$/$cm$^3]$ & $[M_{\odot}]$ & [km] & [kHz] & [ms] &
        [$10^{45}$g\,cm$^2$] & --- \\
        \hline
        \multirow{4}{5em}{$x300 \beta0.5\pi$}
        & 0.560 & 1.175 & 16.22 & 1.220 & 629.6 & 1.735 & 4.57e+03 \\
        & 1.008 & 1.336 & 14.60 & 1.540 & 330.7 & 1.618 & 9.92e+02 \\
        & 1.456 & 1.384 & 13.58 & 1.761 & 246.7 & 1.452 & 4.59e+02 \\
        & 1.904 & 1.394 & 12.86 & 1.930 & 208.7 & 1.306 & 2.84e+02 \\
        \hline
        \multirow{4}{5em}{$x300 \beta1.0\pi$}
        & 0.265 & 1.634 & 23.15 & 0.843 & 958.1 & 4.905 & 5.38e+03 \\
        & 0.494 & 1.885 & 20.73 & 1.080 & 476.4 & 4.600 & 1.04e+03 \\
        & 0.723 & 1.957 & 19.24 & 1.242 & 350.6 & 4.117 & 4.65e+02 \\
        & 0.952 & 1.971 & 18.19 & 1.364 & 295.2 & 3.696 & 2.84e+02 \\
        \hline
        \multirow{4}{5em}{$x300 \beta1.5\pi$}
        & 0.172 & 1.987 & 28.45 & 0.681 & 1209 & 9.006& 5.75e+03 \\
        & 0.326 & 2.306 & 25.43 & 0.879 & 587.8 & 8.466 & 1.05e+03 \\
        & 0.480 & 2.396 & 23.58 & 1.013 & 430.2 & 7.572 & 4.67e+02 \\
        & 0.634 & 2.414 & 22.29 & 1.114 & 361.5 & 6.791 & 2.84e+02 \\
        \hline
        \multirow{4}{5em}{$x400 \beta0.5\pi$}
        & 0.888 & 0.500 & 10.07 & 1.608 & 1026 & 0.2784 & 4.34e+04 \\
        & 2.171 & 0.699 & 8.817 & 2.357 & 276.1 & 0.3063 & 2.58e+03 \\
        & 3.453 & 0.759 & 8.091 & 2.821 & 173.3 & 0.2824 & 8.29e+02 \\
        & 4.735 & 0.779 & 7.600 & 3.162 & 136.1 & 0.2558 & 4.35e+02 \\
        & 6.017 & 0.784 & 7.238 & 3.431 & 117.4 & 0.2326 & 2.84e+02 \\
        \hline
        \multirow{4}{5em}{$x400 \beta1.0\pi$}
        & 0.246 & 0.500 & 15.12 & 0.867 & 4650 & 0.6168 & 4.17e+05 \\
        & 0.937 & 0.951 & 12.78 & 1.570 & 466.0 & 0.8732 & 3.88e+03 \\
        & 1.627 & 1.066 & 11.57 & 1.951 & 258.1 & 0.8108 & 9.45e+02 \\
        & 2.318 & 1.101 & 10.79 & 2.219 & 195.2 & 0.7291 & 4.52e+02 \\
        & 3.008 & 1.109 & 10.23 & 2.426 & 166.0 & 0.6579 & 2.84e+02 \\
        \hline
        \multirow{4}{5em}{$x400 \beta1.5\pi$}
        & 0.123 & 0.500 & 18.93 & 0.617 & 10916 & 0.959 & 1.40e+06 \\
        & 0.593 & 1.147 & 15.79 & 1.255 & 609.1 & 1.605 & 4.49e+03 \\
        & 1.064 & 1.302 & 14.22 & 1.581 & 321.6 & 1.496 & 9.88e+02 \\
        & 1.535 & 1.348 & 13.23 & 1.808 & 240.3 & 1.342 & 4.58e+02 \\
        & 2.005 & 1.358 & 12.53 & 1.981 & 203.3 & 1.208 & 2.84e+02 \\
        \bottomrule
    \end{tabular}
    \label{tab:table-bdm}
\end{table*}

\begin{table*}
    \centering
    \caption{Data for EoSs of fermionic dark matter. The columns are the same as
    in Tab.~\ref{tab:ns}.}
    \ra{1.3}
    \begin{tabular}{cccccccc}
        \toprule
        EoS & $\epsilon_c$ & $M$ & $R$ & $f$ & $\tau$ & $I$ & $\Lambda$ \\
        \hline
        --- & [$10^{15}$g$/$cm$^3]$ & $[M_{\odot}]$ & [km] & [kHz] & [ms] &
        [$10^{45}$g\,cm$^2$] & --- \\
        \hline
        \multirow{4}{6em}{x1~$\phi8$~$\alpha$0.001}
        & 0.213 & 2.269 & 27.62 & 0.768 & 788.7 & 9.802 & 2.05e+03 \\
        & 0.409 & 2.686 & 24.14 & 1.012 & 380.5 & 9.392 & 2.80e+02 \\
        & 0.605 & 2.809 & 22.05 & 1.176 & 286.3 & 8.455 & 1.04e+02 \\
        & 0.801 & 2.833 & 20.63 & 1.300 & 248.6 & 7.590 & 5.67e+01 \\
        \hline
        \multirow{4}{6em}{x1~$\phi10$~$\alpha$0.001}
        & 0.313 & 1.878 & 23.09 & 0.924 & 660.6 & 5.578 & 2.08e+03 \\
        & 0.597 & 2.202 & 20.14 & 1.215 & 322.2 & 5.266 & 3.01e+02 \\
        & 0.881 & 2.298 & 18.39 & 1.413 & 241.8 & 4.722 & 1.13e+02 \\
        & 1.165 & 2.318 & 17.19 & 1.562 & 209.0 & 4.233 & 6.22e+01 \\
        \hline
        \multirow{4}{6em}{x1~$\phi12$~$\alpha$0.001}
        & 0.422 & 1.619 & 20.14 & 1.064 & 579.4 & 3.600 & 2.13e+03 \\
        & 0.800 & 1.882 & 17.53 & 1.398 & 285.4 & 3.351 & 3.25e+02 \\
        & 1.177 & 1.961 & 16.00 & 1.625 & 213.5 & 2.994 & 1.24e+02 \\
        & 1.555 & 1.977 & 14.95 & 1.799 & 183.8 & 2.680 & 6.85e+01 \\
        \hline
        \multirow{4}{6em}{x2~$\phi8$~$\alpha$0.001}
        & 0.250& 0.500 & 15.11 & 0.878 & 4508 & 0.606 & 3.99e+05 \\
        & 1.097 & 1.122 & 12.40 & 1.752 & 299.9 & 1.023 & 1.23e+03 \\
        & 1.944 & 1.296 & 11.05 & 2.218 & 164.8 & 0.987 & 2.14e+02 \\
        & 2.792 & 1.349 & 10.19 & 2.541 & 128.9 & 0.899 & 8.58e+01 \\
        & 3.639 & 1.360 & 9.58 & 2.788 & 113.9 & 0.814 & 4.83e+01 \\
        \hline
        \multirow{4}{6em}{x2~$\phi10$~$\alpha$0.001}
        & 0.506 & 0.500 & 12.05 & 1.240 & 1928 & 0.388 & 1.14e+05 \\
        & 1.760 & 0.925 & 10.01 & 2.204 & 229.4 & 0.548 & 1.04e+03 \\
        & 3.014 & 1.050 & 8.965 & 2.749 & 132.7 & 0.523 & 2.09e+02 \\
        & 4.268 & 1.090 & 8.284 & 3.133 & 105.0 & 0.476 & 8.75e+01 \\
        & 5.522 & 1.098 & 7.798 & 3.429 & 92.99 & 0.431 & 5.01e+01 \\
        \hline
        \multirow{4}{6em}{x2~$\phi12$~$\alpha$0.001}
        & 0.914 & 0.500 & 9.989 & 1.652 & 963.2 & 0.2691 & 3.93e+04 \\
        & 2.613 & 0.795 & 8.416 & 2.665 & 182.2 & 0.3319 & 8.82e+02 \\
        & 4.311 & 0.888 & 7.573 & 3.271 & 111.3 & 0.3134 & 2.03e+02 \\
        & 6.009 & 0.918 & 7.017 & 3.706 & 89.14 & 0.2854 & 8.91e+01 \\
        & 7.708 & 0.924 & 6.617 & 4.046 & 79.15 & 0.2594 & 5.21e+01 \\
        \bottomrule
    \end{tabular}
    \label{tab:table}
\end{table*}

\newpage
\twocolumngrid

\end{document}